\pdfoutput=1
\documentclass[11pt,letterpaper]{article}
\usepackage{jheppub}

\DeclareMathOperator{\Li}{Li}
\newcommand{\ket}[1]{\langle #1 \rangle}

\newcommand{\be}{\begin{equation}}
\newcommand{\ee}{\end{equation}}
\newcommand{\bea}{\begin{eqnarray}}
\newcommand{\eea}{\end{eqnarray}}
\newcommand{\ud}{\mathrm d}

\newcommand{\reef}[1]{(\ref{#1})}
\newcommand{\gn}[1]{\Gamma\left(#1\right)}

\newcommand{\s}[1]{\Delta^+_{#1}}
\newcommand{\df}[1]{\Delta^-_{#1}}
\newcommand{\dij}{\hat \partial_{ij}}

\title{Mellin Amplitudes for Dual Conformal Integrals}

\author[a]{Miguel F. Paulos,}
\author[b]{Marcus Spradlin}
\author[b]{and Anastasia Volovich}

\affiliation[a]{Laboratoire de Physique Th\'eorique et Hautes Energies, CNRS UMR 7589, Universit\'e Pierre et Marie Curie, 4 place Jussieu, 75252 Paris Cedex 05, France}
\affiliation[b]{Department of Physics, Brown University, Box 1843, Providence, RI 02912-1843, USA}

\emailAdd{mpaulos@lpthe.jussieu.fr}
\emailAdd{marcus\_spradlin@brown.edu}
\emailAdd{anastasia\_volovich@brown.edu}

\abstract{Motivated by recent work on the utility of Mellin space
for representing conformal correlators in $AdS$/CFT, we study its suitability for representing dual conformal integrals of the type which appear in perturbative scattering amplitudes in super-Yang-Mills theory.
We discuss Feynman-like rules for writing Mellin amplitudes for a large
class of integrals in any dimension, and
find explicit representations for several familiar toy integrals.
However we show that the power
of Mellin space
is that it provides simple representations even for fully massive integrals,
which except for the single
case of the 4-mass box
have not yet been computed by any available technology.
Mellin space is also useful for exhibiting differential relations between
various multi-loop integrals, and
we show that certain higher-loop integrals
may be written
as integral operators acting on the fully massive
scalar $n$-gon in
$n$ dimensions, whose Mellin amplitude is exactly 1.
Our chief example is a
very simple formula expressing the 6-mass
double box as a single integral of the 6-mass
scalar hexagon in 6 dimensions.
}

\begin{document}

\maketitle

\section{Introduction}

The interplay between conceptual and technical advances has always been an important catalyst
for progress in theoretical physics.  A spectacular example
of this has been the
development of our understanding of
$\mathcal{N}=4$ supersymmetric Yang-Mills (SYM) theory~\cite{Brink:1976bc}---the
exemplar of four-dimensional quantum field theories---especially since the discovery
of the $AdS$/CFT correspondence~\cite{Aharony:1999ti} almost 15 years ago.
In particular, two related aspects of planar SYM theory which have generated considerable
attention within the past several years include
the application of powerful integrability techniques~\cite{INTReview} for determining operator
dimensions and the discovery of remarkable mathematical structure in perturbative scattering
amplitudes~\cite{SCATReview}.

More recently
some attention has focused on the important
problem of understanding better the structure of correlation functions in SYM theory, which
at strong coupling may be computed via $AdS$/CFT using Witten diagrams~\cite{Witten:1998qj}.
Sometimes, both conceptual and technical progress can be aided by `using the right language'
(or set of variables), as
dramatically evidenced for example by the use of spinor helicity and momentum twistor
variables for scattering amplitudes.  Motivated in part by appreciation of this lesson,
it has been suggested~\cite{Fitzpatrick:2011ia} that the `right' place to study $AdS$/CFT
correlation functions is not in position space but rather in Mellin space,
the benefits of which for correlation functions in general CFTs were pointed out
in the pioneering work of Mack~\cite{Mack:2009mi}.

For the purpose of studying $AdS$/CFT correlation functions Mellin space is both
healthy and great tasting---it allows for a dramatic simplification of otherwise
intractable computations (see for example~\cite{Penedones:2010ue,Paulos:2011ie,Nandan:2011wc}),
while at the same time providing
a definition for a dual bulk S-matrix via the vanishing curvature limit of the
$AdS$/CFT correspondence~\cite{Penedones:2010ue,Fitzpatrick:2011dm,Fitzpatrick:2011ia}. Mellin amplitudes also make the physics of correlators transparent, by showing in a simple way their conformal block decomposition, and their associated OPE coefficients. This is analogous to the way in which momentum space clarifies the physics of weakly coupled field theories. It seems therefore that a strong case can be made that {\em Mellin amplitudes} are the right object to consider in any conformally invariant setting.

Motivated by the promise of this approach, our goal in this paper is to carry out a first
study of the suitability of Mellin space as a language for the
the weak-coupling expansion of scattering amplitudes SYM theory.
Let us emphasize right away that we are interested here in the {\it boundary} flat space S-matrix,
rather than the {\it bulk} S-matrix for supergravity or string theory in the flat space limit
of $AdS$.
The latter has been studied since the earliest days of $AdS$/CFT
(see for example~\cite{Polchinski:1999ry,Susskind:1998vk,Giddings:1999qu})
and may be computed as a certain limit of SYM theory correlation functions
(see~\cite{Okuda:2010ym,Penedones:2010ue,Fitzpatrick:2011hu,Raju:2012zr} for recent work).
The former may be computed at strong coupling
via $AdS$/CFT by introducing a probe brane to provide the
gluon degrees of freedom~\cite{Alday:2007hr}, but in this paper our interest lies in the
perturbative expansion at weak coupling.

Although we consider examples of integrals which may appear in various field theories,
SYM theory is apparently unique amongst four-dimensional field theories in that
in that {\it all} of its scattering amplitudes are amenable
to Mellin representations of the type
we discuss since they all possess dual conformal invariance~\cite{Drummond:2007cf,Drummond:2007au,Drummond:2008vq}.
This symmetry of SYM theory was first noticed in some examples~\cite{Drummond:2006rz} and
proven to be a property of the integrand for general amplitudes to all loop order
in~\cite{ArkaniHamed:2010gh}.

One of our motivations for seeking a new language for loop integrals in flat space is that
despite remarkable recent advances, actually carrying out multi-loop integrals remains
a very challenging task for which there is no practical general algorithm
(see~\cite{SmirnovBook} and references therein for some of the most modern magic).
This stands in stark contrast to the situation for the planar integr{\it \!and} of SYM
theory,
which is amenable to powerful generalized unitarity techniques~\cite{Bern:1994cg,Bern:1994zx,Bern:2007ct}
and which can in principle be computed for any desired process via the recursion
described
in~\cite{ArkaniHamed:2010kv} (see also~\cite{CaronHuot:2010zt,Boels:2010nw}).
Given the current relative simplicity of computing integrands but the relative difficulty of
computing integrals it is natural to wonder whether there exists some kind of `stepping stone'
in between these two quantities.  Initially we should not necessarily require this stepping stone to exist
for arbitrary
amplitudes in any random field theory, only for the very special amplitudes of planar SYM theory.
However
we should require it to be completely canonical---both the integrand
and the integrated amplitudes of SYM theory are mathematically well-defined objects which look identical
to us today as they will to an alien civilization a billion years in the future, and the same should
be true of any good stepping stone.

One important example of something halfway between an integrand and its
integral, for those integrals which can be expressed in terms of a certain class of generalized
polylogarithm functions, is the `symbol' described for
example in~\cite{Goncharov:2002,Goncharov:1994,Goncharov:1998} and
first used for SYM theory amplitudes in~\cite{Goncharov:2010jf}.
Symbol technology has proven very useful in several applications
(see for example~\cite{DelDuca:2011ne,Dixon:2011ng,DelDuca:2011jm,DelDuca:2011wh,
Spradlin:2011wp,Dixon:2011pw,Dixon:2011nj,Prygarin:2011gd,Heslop:2011hv,Duhr:2011zq,Duhr:2012fh,CaronHuot:2011kk,Gaiotto:2011dt,Brandhuber:2012vm}),
but it appears that sufficiently complicated amplitudes even in SYM theory involve
elliptic functions (of a type familiar in the QCD literature,
see for example~\cite{Laporta:2004rb}) which are outside the class
treatable by current symbol technology.

In this paper we propose that Mellin space representations of the type
recently employed for $AdS$ correlation functions might provide useful
also for flat space dual conformally invariant amplitudes. After quickly introducing the class of integrals under consideration in section~\ref{settingupsec} and the definition of the Mellin amplitude in section~\ref{mellinsec},
we explore aspects
of this proposal via
several examples in the subsequent sections.  In particular, we propose in section~\ref{feynmansec} that there are simple Feynman-like rules for directly obtaining the Mellin amplitudes corresponding to dual conformal integrals with trivial numerator factors. These rules are considerably simpler than the ones proposed in $AdS$/CFT, and we show how the former seemingly derive from the latter. In practice they amount to thinking of the position space diagram as a diagram in {\em Mellin momentum space}; in this way internal lines map to propagator factors and contact interactions have Mellin amplitude equal to one. This remarkable fact has important consequences for deriving differential equations for dual conformal integrals, and also allows one to immediately {\em solve} such equations, thereby reducing the computation of certain higher-loop diagrams to simple integrals of one-loop $n$-gon diagrams.

In sections~\ref{pentasec} and~\ref{hexsec}
we investigate the Mellin representation of conformal integrals with numerator factors, and in particular the pure chiral integrals studied extensively in~\cite{ArkaniHamed:2010gh}. We find several interesting features: firstly, the magic numerator factors appearing in the chiral pentagon and hexagon integrals make some of the Mellin integrals ``collapse'' onto boundary poles. We explain how this works in detail in a simple example, and then use this property to derive a representation of the chiral pentagon in terms of derivatives of hypergeometric functions. Next we show how for the chiral hexagon, double numerator factors translate into second-order differential operators in the Mellin representation.  Such differential equations have been studied in~\cite{Drummond:2010cz,Dixon:2011ng}, and our results make it clear that they easily generalizes to more complicated examples.

We finish this paper with a short discussion, followed by various appendices containing additional technical results and details.

\section{Setting up}
\label{settingupsec}

\subsection{Dual conformal integrals in SYM theory}
\label{sasec}

Here we provide a quick introduction to the important
features of the integrals which appear in SYM theory loop amplitudes.
The most important property, which has been proven to hold
to all orders in perturbation theory for all amplitudes, is that
the integrand is invariant\footnote{Strictly speaking it is
only
covariant, but it is rendered invariant after dividing by the
tree-level MHV superamplitude, which we follow standard convention in doing.}
under dual conformal
transformations.
We remind the reader that dual conformal transformations~\cite{Drummond:2008vq}
are nothing
but ordinary conformal transformations on the dual variables $x_1,\ldots,x_n$
related to the momenta of the $n$  scattering particles by
\begin{equation}
p_i = x_i - x_{i + 1}, \qquad x_{ij} \equiv (x_i - x_j)^2
\end{equation}
where all subscripts are taken mod $n$.

\begin{figure}
\begin{center}
\begin{picture}(300,100)
\put(-5,15){\includegraphics[width=70pt]{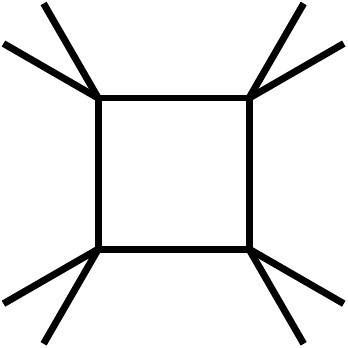}}
\put(115,15){\includegraphics[width=70pt]{1loopmassive.pdf}}
\put(235,15){\includegraphics[width=70pt]{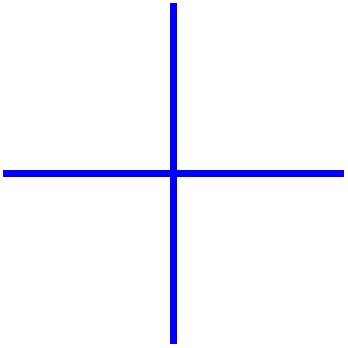}}
\put(3,8){\makebox(0,0){$1$}}
\put(-10,20){\makebox(0,0){$2$}}
\put(-10,82){\makebox(0,0){$3$}}
\put(3,92){\makebox(0,0){$4$}}
\put(57,8){\makebox(0,0){$8$}}
\put(71,20){\makebox(0,0){$7$}}
\put(58,92){\makebox(0,0){$5$}}
\put(71,82){\makebox(0,0){$6$}}
\put(30,25){\makebox(0,0){$x_1$}}
\put(5,50){\makebox(0,0){$x_3$}}
\put(30,75){\makebox(0,0){$x_5$}}
\put(55,50){\makebox(0,0){$x_7$}}
\put(150,25){\makebox(0,0){$x_1$}}
\put(125,50){\makebox(0,0){$x_2$}}
\put(150,75){\makebox(0,0){$x_3$}}
\put(175,50){\makebox(0,0){$x_4$}}
\put(270,5){\makebox(0,0){$x_1$}}
\put(225,50){\makebox(0,0){$x_2$}}
\put(270,95){\makebox(0,0){$x_3$}}
\put(315,50){\makebox(0,0){$x_4$}}
\put(30,-15){\makebox(0,0){(a)}}
\put(150,-15){\makebox(0,0){(b)}}
\put(270,-15){\makebox(0,0){(c)}}
\end{picture}
\end{center}
\caption{The one-loop 8-point four-mass integral, labeled according to
the usual amplitude convention in (a) and according to our streamlined
notation in (b).  In (a) it is implicit in the notation that each $x_i$ should be null-separated
from its neighbors $x_{i-1}$ and $x_{i+1}$.  In contrast the $x_i$ in (b) and (c) are arbitrary, and
(a) is recovered by a simple relabeling. This integral corresponds in the dual (Mellin momentum)
space to a tree-level contact
interaction (c).}
\label{fig:1loopmassive}
\end{figure}

We take this opportunity to immediately break with the standard conventions
of the amplitude community in order to streamline the notation for this
paper.
Instead of using dual variables $x_i$ which implicitly are null separated from their neighbors
({\em i.e.} $(x_i-x_{i+1})^2 = 0$), we will only
use as many dual variables as external faces in any diagram under
consideration, and furthermore let them initially take generic values.  For example consider the four-mass\footnote{We
remind the reader unfamiliar with this terminology that the label `four-mass'
is used because the sum of the external momenta entering each of the four corners
of the box is non-null.  There is no $+ m^2$ in any propagator and there is
no breaking the conformal (or dual conformal) symmetry of SYM theory.}
integral shown
in Fig.~\ref{fig:1loopmassive}.  In the amplitude convention this diagram would be labeled with eight $x$'s, but since the value of the integral
only depends on the four $x$'s shown in (a), we can economize the notation
by relabeling the diagram as shown in (b).
The value of this integral is then
\begin{equation}
\label{eq:1loopmassive}
{\hbox{\lower 12pt\hbox{
\begin{picture}(30,30)
\put(0,0){\includegraphics[width=30pt]{1loopmassive.pdf}}
\end{picture}
}}}=
\int \frac{d^4 x}{i \pi^2} \frac{x_{13}^2 x_{24}^2}{(x-x_1)^2(x-x_2)^2
(x-x_3)^2 (x-x_4)^2}.
\end{equation}
Note, importantly, the inclusion of the overall factor $x_{13}^2 x_{24}^2$
into the definition of this integral.  This factor, which is
required for dual conformal invariance,
is the first reminder that the integrals under consideration here are not
exactly those of scalar $\phi^4$ theory, though in many cases
they are very closely related.

Of course~(\ref{eq:1loopmassive}) is manifestly
the same (again, up to the factor $x_{13}^2 x_{24}^2$) integral
which computes the {\it tree-level} position
space CFT correlation function
$\langle \mathcal{O}(x_1) \mathcal{O}(x_2)
\mathcal{O}(x_3) \mathcal{O}(x_4)\rangle$ of four
operators with dimension $\Delta = 1$ interacting via a four-point
contact interaction, as shown in Fig.~\ref{fig:1loopmassive}(c).
Henceforth we will always draw the dual diagram (c) in blue directly
on top of the corresponding integral (b) in order to save space.

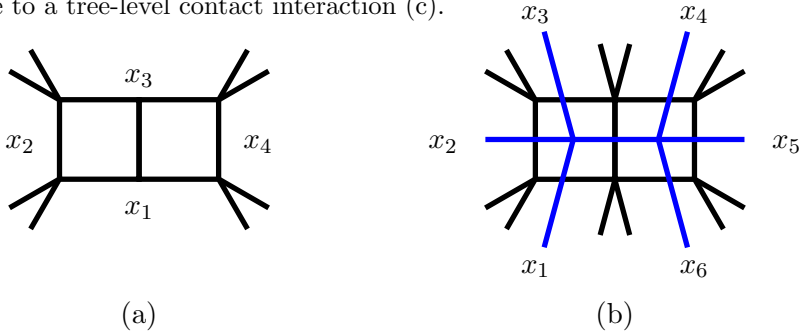
\begin{figure}
\begin{center}
\begin{picture}(300,85)
\put(10,17){\includegraphics[width=100pt]{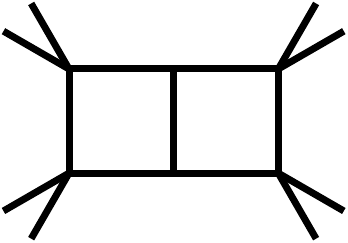}}
\put(190,10){\includegraphics[width=100pt]{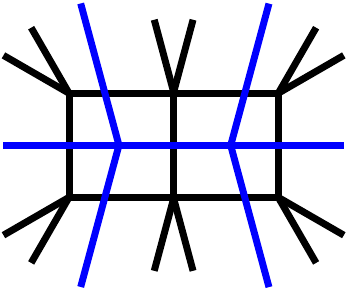}}
\put(60,-15){\makebox(0,0){(a)}}
\put(240,-15){\makebox(0,0){(b)}}
\put(60,25){\makebox(0,0){$x_1$}}
\put(15,50){\makebox(0,0){$x_2$}}
\put(60,75){\makebox(0,0){$x_3$}}
\put(105,50){\makebox(0,0){$x_4$}}
\put(175,50){\makebox(0,0){$x_2$}}
\put(305,50){\makebox(0,0){$x_5$}}
\put(210,3){\makebox(0,0){$x_1$}}
\put(270,3){\makebox(0,0){$x_6$}}
\put(210,99){\makebox(0,0){$x_3$}}
\put(270,99){\makebox(0,0){$x_4$}}
\end{picture}
\end{center}
\caption{The two-loop four-mass double box
integral (a) is a particular limit of the `fully massive' double box (b),
computed in Mellin space as an exchange diagram contribution to a
tree-level 6-point correlation function (in blue). The integral (a) is recovered
from (b) by taking the limit $x_4 \to x_3$, $x_6 \to x_1$ and then relabeling $x_5 \to x_4$.  We define
the integral (b) to include the overall factor $x_{14}^2 x_{25}^2 x_{36}^2$ in order to provide dual conformal invariance.
This reduces to the factor $x_{13}^4 x_{24}^2$ for integral (a).}
\label{fig:2loopmassive}
\end{figure}

In our approach it is most natural to always begin with the
{\it fully massive} version of any integral under consideration, and then to
recover other versions of the integral by taking appropriate limits.  For example,
the two-loop four-mass diagram shown in Fig.~\ref{fig:2loopmassive}(a) is
a perfectly nice finite and dual conformal invariant integral,
but we represent
it as a
limit of the fully massive integral as shown in
Fig.~\ref{fig:2loopmassive}(b) and given by
\begin{equation}
\label{eq:2loopmassive}
{\hbox{\lower 16pt\hbox{
\begin{picture}(50,37)
\put(0,0){\includegraphics[width=50pt]{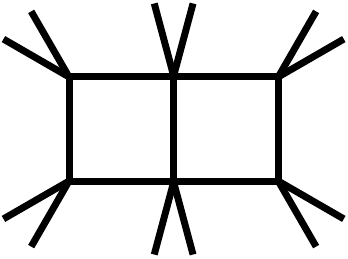}}
\end{picture}
}}}=
\int \frac{d^4 x_a}{i \pi^2} \frac{d^4 x_b}{i \pi^2} \frac{x_{14}^2 x_{25}^2 x_{36}^2}{x_{a1}^2 x_{a2}^2 x_{a3}^2 x_{ab}^2
x_{b4}^2 x_{b5}^2 x_{b6}^2}.
\end{equation}

The fact that fully massive integrals are often the simplest to work with
in Mellin space is one of its most attractive features, since it is opposite
to the experience of amplitudeologists to whom more massive
integrals are necessarily more complicated.  For example,
while it remains an open challenge to evaluate the integral in Fig.~\ref{fig:2loopmassive}(b)
at just two loops
(it is believed to involve
elliptic functions), even the $L$-loop generalization of the integral in
(a) was fully evaluated long ago~\cite{Usyukina:1992jd} in terms of
standard polylogarithm functions,
\begin{equation}
\label{eq:lloopmassive}
{\hbox{\lower 16pt\hbox{
\begin{picture}(75,40)
\put(0,0){\includegraphics[width=75pt]{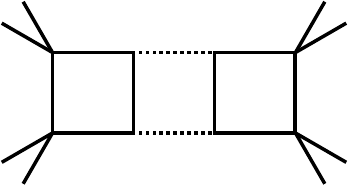}}
\end{picture}
}}}=
- \frac{1}{L! \lambda} \sum_{j=L} \frac{(-1)^j j! \log^{2L-j}(v/u)}{(j-L)! (2L-j)!}
\left[\Li_j\left(-\frac{1}{\rho u}\right) - \Li_j(-\rho v) \right]
\end{equation}
where
\begin{equation}
\lambda = \sqrt{(1-u-v)^2-4 u v}, \qquad \rho=  \frac{2}{1-u-v+\lambda}, \qquad
u = \frac{x_{12}^2 x_{34}^2}{x_{13}^2 x_{24}^2}, \qquad
v = \frac{x_{14}^2 x_{23}^2}{x_{13}^2 x_{24}^2}.
\end{equation}
We will see below in section~\ref{feynmanrules} that it is trivial
to write down the $L$-loop generalization of the fully massive
integral in Fig.~\ref{fig:2loopmassive}(b) in Mellin space,
from which~(\ref{eq:1loopmassive}) would follow as a special case.
Let us however temper our enthusiasm (slightly) by pointing out that taking
such limits of interest is often but not always a trivial task in Mellin space, as
we discuss below in section~\ref{sec:massless}.

\begin{figure}
\begin{center}
\begin{picture}(200,72)
\put(0,0){\includegraphics[width=200pt]{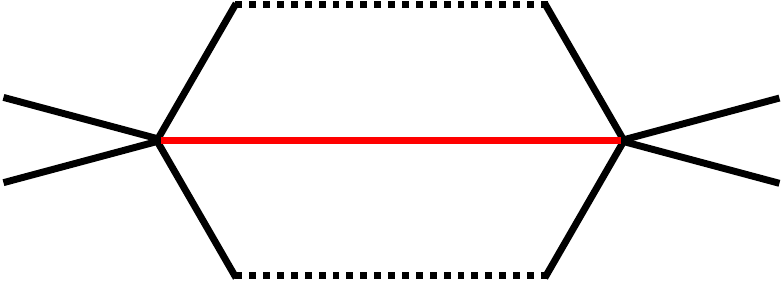}}
\put(30,56){\makebox(0,0){$x_{i+1}$}}
\put(30,16){\makebox(0,0){$x_i$}}
\put(170,56){\makebox(0,0){$x_j$}}
\put(170,16){\makebox(0,0){$x_{j+1}$}}
\put(100,53){\makebox(0,0){$x$}}
\end{picture}
\end{center}
\label{fig:magicnumerator}
\caption{`Magic' numerator factors are denoted graphically by a red line crossing an internal face.}
\end{figure}
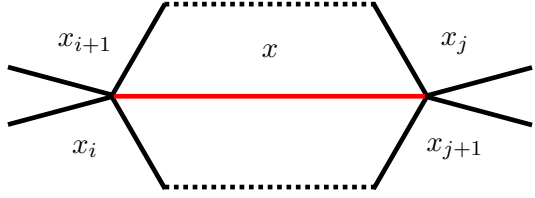

The correspondence between the integrals
appearing in SYM theory and in $\phi^4$ theory only
holds for the simplest diagrams.  General integrands in SYM theory
have non-trivial numerator factors.
A particularly nice collection of such integrals are those
involving
chiral numerator factors of the type discussed
extensively in~\cite{Drummond:2010mb,ArkaniHamed:2010kv}.
Denoted graphically by a red line crossing some internal face (see Fig.~\ref{fig:magicnumerator}),
the corresponding numerator factor is proportional to\footnote{By convention
chiral integrals are normalized so their nonzero leading singularity is 1.}
the quantity $(x - y)^2$, where $y$ is
a solution to the leading singularity equations
\begin{equation}
(y-x_i)^2 = (y-x_{i+1})^2 = (y-x_j)^2 = (y-x_{j+1})^2 = 0.
\end{equation}
For given $x$'s these equations have two different solutions for $y$;
the corresponding two numerator factors were denoted by squiggly and
dashed red lines in~\cite{ArkaniHamed:2010kv}.
The utility of these perhaps strange-looking numerator factors is
precisely that by killing one of the leading singularities associated
with the loop integration variable $x$ they allow one to express
various integrands in SYM theory very compactly (especially for example MHV amplitudes,
as those are maximally chiral).

\begin{figure}
\begin{center}
\begin{picture}(300,120)
\put(0,10){\includegraphics[width=100pt]{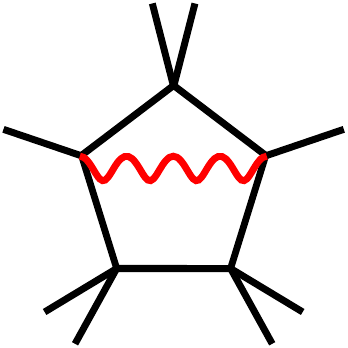}}
\put(200,10){\includegraphics[width=100pt]{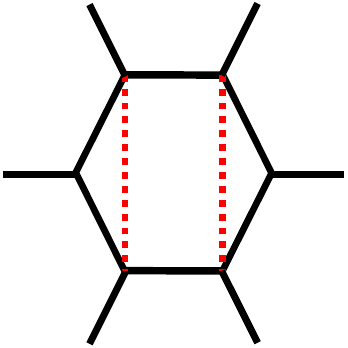}}
\put(15,45){\makebox(0,0){$x_1$}}
\put(25,85){\makebox(0,0){$x_2$}}
\put(75,85){\makebox(0,0){$x_3$}}
\put(85,45){\makebox(0,0){$x_4$}}
\put(50,20){\makebox(0,0){$x_5$}}
\put(250,20){\makebox(0,0){$x_1$}}
\put(250,100){\makebox(0,0){$x_4$}}
\put(215,40){\makebox(0,0){$x_2$}}
\put(215,80){\makebox(0,0){$x_3$}}
\put(285,80){\makebox(0,0){$x_5$}}
\put(285,40){\makebox(0,0){$x_6$}}
\put(50,-15){\makebox(0,0){(a)}}
\put(250,-15){\makebox(0,0){(b)}}
\end{picture}
\end{center}
\caption{The chiral pentagon (a) and hexagon (b) integrals under consideration in this paper.}
\label{fig:penthex}
\end{figure}
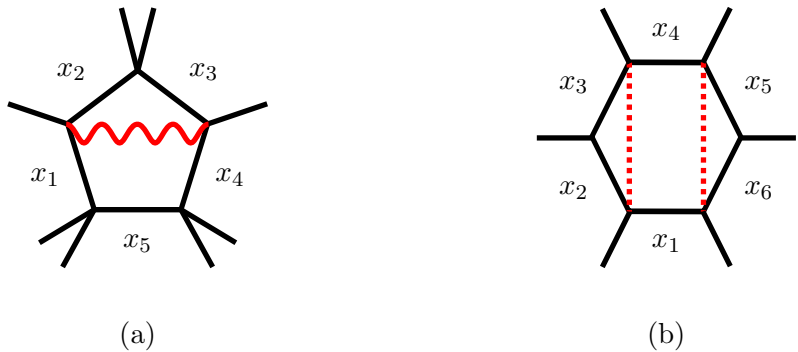

In this paper we will study in detail one-loop chiral integrals with one and two numerator factors,
two particularly simple examples of which are
shown in Fig.~\ref{fig:penthex}.
Here we have drawn the pentagon with two massless corners and the hexagon with all massless corners.
We will actually begin both cases by considering an arbitrary (fully massive) one-loop $n$-gon integral with one or two
numerators and then take appropriate limits to reach these two special cases.
Our interest in them in particular stems from the fact that these are the configurations
in which the integrals enter one-loop MHV and NMHV amplitudes in SYM
theory~\cite{ArkaniHamed:2010gh}.
In these limits the numerator factors simplify and the integrands can
be written rather simply in terms of momentum twistors~\cite{Hodges:2009hk}.  For the pentagon we need at least 8 legs to provide
the 3 massive corners.  For later use let us choose to label the legs by their momentum twistor variables $Z_i$, $i=1,\ldots,8$ in this case as
\begin{equation}
\label{eq:pentagonlabeling}
{\hbox{\lower 35pt\hbox{
\begin{picture}(80,80)
\put(10,10){\includegraphics[width=60pt]{pentagon.pdf}}
\put(18,3){\makebox(0,0){$1$}}
\put(8,13){\makebox(0,0){$2$}}
\put(3,50){\makebox(0,0){$3$}}
\put(62,3){\makebox(0,0){$8$}}
\put(72,13){\makebox(0,0){$7$}}
\put(48,75){\makebox(0,0){$5$}}
\put(32,75){\makebox(0,0){$4$}}
\put(77,50){\makebox(0,0){$6$}}
\end{picture}
}}}=
\int_{AB} \frac{\ket{AB (234) \cap (567)} \ket{3681}}{\ket{AB23} \ket{AB34} \ket{AB56} \ket{AB67} \ket{AB81}},
\end{equation}
while the massless hexagon is simply
\begin{equation}
\label{eq:hexagon}
{\hbox{\lower 35pt\hbox{
\begin{picture}(80,80)
\put(0,10){\includegraphics[width=60pt]{hexagon.pdf}}
\put(12,3){\makebox(0,0){$1$}}
\put(46,3){\makebox(0,0){$6$}}
\put(12,76){\makebox(0,0){$3$}}
\put(46,76){\makebox(0,0){$4$}}
\put(-8,40){\makebox(0,0){$2$}}
\put(67,40){\makebox(0,0){$5$}}
\end{picture}
}}}=
\int_{AB} \frac{\ket{AB13} \ket{AB46} \ket{5612} \ket{2345}}{\ket{AB12} \ket{AB23} \ket{AB34} \ket{AB45} \ket{AB56} \ket{AB61}}.
\end{equation}
These two integrals were evaluated explicitly in~\cite{Drummond:2010cz}
with the results
\begin{eqnarray}
\label{eq:pentagonresult}
{\hbox{\lower 12pt\hbox{
\begin{picture}(30,30)
\put(0,0){\includegraphics[width=30pt]{pentagon.pdf}}
\end{picture}
}}}&=&
\Li_2(1-u_{13}) + \Li_2(1-u_{35}) + \Li_2(1-u_{14}) \nonumber \\
&& \qquad  - \Li_2(1-u_{13} u_{35}) - \Li_2(1-u_{13} u_{14}) + \log(u_{35}) \log(u_{14}),
\\
\label{eq:hexagonresult}
{\hbox{\lower 12pt\hbox{
\begin{picture}(30,30)
\put(0,0){\includegraphics[width=30pt]{hexagon.pdf}}
\end{picture}
}}}&=&
\Li_2(1-u_{14}) + \Li_2(1-u_{25}) + \Li_2(1-u_{36}) + \log(u_{25}) \log(u_{36}) - \frac{\pi^2}{3},
\end{eqnarray}
where
\begin{equation}
\label{eq:uij}
u_{ij} = \frac{x_{i,j+1}^2 x_{i+1,j}^2}{x_{i,j}^2 x_{i+1,j+1}^2}.
\end{equation}

Since the formalism we will employ works in arbitrary dimension we will also encounter higher dimensional integrals, including
the six-dimensional scalar hexagon
\begin{equation}
{\hbox{\lower 27pt\hbox{
\begin{picture}(60,60)
\put(0,0){\includegraphics[width=60pt]{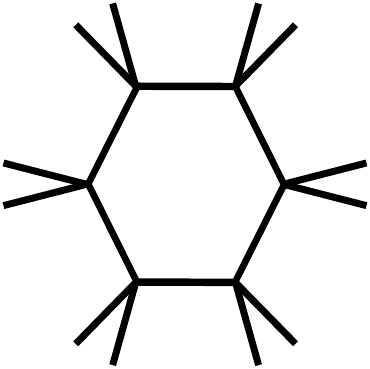}}
\put(30,30){\makebox(0,0){$d{=}6$}}
\end{picture}
}}}
= \int \frac{d^6 x}{i \pi^3} \frac{x_{14}^2 x_{25}^2 x_{36}^2}{(x-x_1)^2(x-x_2)^2(x-x_3)^2(x-x_4)^2(x-x_5)^2(x-x_6)^2}.
\end{equation}
This integral
has been evaluated in special cases including zero~\cite{DelDuca:2011ne,Dixon:2011ng}, one~\cite{DelDuca:2011jm} and
three~\cite{DelDuca:2011wh} masses, but not yet for completely general $x_i$ (though its symbol is known~\cite{Spradlin:2011wp}).
The importance of this integral for SYM theory scattering amplitudes, and its relation to the integral~\reef{eq:hexagon}
has been explored and emphasized in~\cite{Dixon:2011ng}.

\subsection{Ambient space formalism}
\label{ambientspace}

The calculation of conformally invariant integrals is conveniently performed in the embedding or ambient space formalism, which goes back to Dirac~\cite{Dirac:1936fq}. For a more recent reference with several details on the four-dimensional case see~\cite{Weinberg:2010fx}, and for some interesting recent applications to CFTs see~\cite{Costa:2011mg,Costa:2011dw}. The basic idea is that we can parameterize $d$-dimensional Minkowski space by projective light-cone coordinates in $d+2$ dimensions. The $SO(d,2)$ invariance group in $d+2$ dimensions is precisely the same as the conformal group in $d$ dimensions. In this way conformal transformations on the coordinates of the $d$-dimensional space become simple rotations of the $d+2$-dimensional coordinate vectors.

More concretely, consider null vectors in $d+2$ dimensions which we will invariably denote by capital letters $P,Q,Y$ and so on.
That these are projective null vectors means we have
	\be
	P^M P_M=-P^+ P^-+P^\mu P_\mu=0, \qquad P\simeq \lambda P
	\ee
where $M, \mu$ are $d+2$ and $d$-dimensional indices respectively.

To obtain coordinates in $d$-dimensional Minkowski space we need to define a reference vector, call it $I$, thereby explicitly breaking conformal invariance. Then the vector $\hat P^M=P^M/(-P\cdot I)$ parameterizes $d$-dimensional flat space. In practice we make the convenient choice
	\be
	\frac{P^M}{-P\cdot I}=\sqrt{2}\, (1,x^2, x^\mu),
	\ee
where we have used light-cone coordinates for the first two entries of the vector. The above choice is equivalent to setting $-P\cdot I=\frac{\sqrt{2}}2 P^+$. By contracting two independent $P$ vectors we obtain
	\be
	P_{ij}\equiv \frac{(-P_i \cdot P_j)}{(-P_i \cdot I)(-P_j \cdot I)}=(x_i-x_j)^2.
	\ee
In practice we will drop factors $P_i\cdot I$ throughout this paper. These can always be recovered by demanding that amplitudes should be invariant under $P_i\to \alpha P_i$, but in any case they always cancel out of any conformally invariant expression.

The embedding formalism is of course well-defined in any dimension. However it is important to notice that using null vectors for describing positions allows us to use various spinor-helicity formalisms for particular cases. In $d=4$ for instance one can use the equivalence $SO(4,2)\simeq SU(4)$ to rewrite six-dimensional vectors $P^M$ as bi-twistors $P^{[AB]}$, where $A,B$ are $SU(4)$ valued. Since these vectors are null, the corresponding bitwistors are simple---that is, they can be written as a product of two $SU(4)$ valued spinors\footnote{These are of course nothing but Hodges' momentum twistors~\cite{Hodges:2009hk}.}
	\be
\label{eq:mt}
	P^M \to P^{AB}= \epsilon^{\alpha \beta} Z_\alpha^A Z_\beta^B= Z_1^{[A} Z_2^{B]},
	\ee
which establishes a map from a $d$-dimensional vector $x^\mu$ onto a line in twistor space.  However for most of our calculations it will suffice for us to use the embedding formalism, which has the advantage of being applicable in any dimension.

\section{Definition and properties of Mellin amplitudes}
\label{mellinsec}

In a conformal field theory the two-point function is uniquely determined (up to an irrelevant normalization constant) to be
	\be
	\left \langle \phi_{\Delta}(P_1) \phi_{\Delta}(P_2) \right \rangle=\frac{1}{(P_{12})^{\Delta}}. \label{2pt}
	\ee
Two-point functions of fields with differing $\Delta$ are identically zero. The parameter $\Delta$ is known as the conformal dimension of the field $\phi$. The three-point function is also uniquely determined by conformal symmetry up to a coupling constant,
	\be
	\left \langle \phi_{\Delta_1}(P_1) \phi_{\Delta_3}(P_2) \, \phi_{\Delta_3}(P_3)\right \rangle=C_{\Delta_1 \Delta_2 \Delta_3}\, \prod_{i<j} P_{ij}^{-\delta_{ij}}
	\ee
with $\delta_{12}=\frac{\Delta_1+\Delta_2-\Delta_3}2,$ and cyclic permutations thereof.
	Higher-point correlation functions are determined up to an arbitrary function of cross-ratios, which are homogeneous combinations of internal products of $P_i$ vectors. For instance, a conformally invariant four-point function can be written in the form
	\be
	\left \langle
	\phi_{\Delta_1}(P_1) \phi_{\Delta_2}(P_2)
	\phi_{\Delta_3}(P_3) \phi_{\Delta_4}(P_4)
	\right \rangle
=\frac{(P_{24}/P_{14})^{\frac{\Delta_1-\Delta_2}2} \, (P_{14}/P_{13})^{\frac{\Delta_3-\Delta_4}2}}{(P_{12})^{\frac{\Delta_1+\Delta_2}2}\,(P_{34})^{\frac{\Delta_3+\Delta_4}2}}\, F(u,v)
	\ee
where $u$ and $v$ are the cross-ratios
	\be
	u=\frac{P_{12} P_{34}}{P_{13} P_{24}}, \qquad v=\frac{P_{14}P_{23}}{P_{13} P_{24}}.
	\ee

Typically the function $F(u,v)$ is a very complicated object. This is not unexpected, since we know that correlation functions in position space do not usually have a simple structure, even for weakly coupled field theories. However, in that case we know what we should do: instead of working in position space we Fourier transform to momentum space. There the analytic properties of the amplitude are very simple, and their physical meaning is clear. At tree-level one sees simple poles corresponding to single-particle states, and branch cuts at loop level corresponding to multi-particle exchange.
For a generic conformal field theory, the Fourier transform is not so useful. This is simply because typical CFTs are strongly coupled, and as such we do not expect that going into a basis of approximately free momentum eigenstates should help. Perhaps the easiest way to see this is to simply consider the Fourier transform of the two-point function~\reef{2pt}. This behaves as $\simeq p^{-2(d-\Delta)}$ and so we see that even the two-point function already shows branch cuts for generic $\Delta$.

The situation therefore might seem hopeless, but the large symmetry of conformal field theory comes to our rescue. Following Mack~\cite{Mack:2009mi} we introduce the Mellin amplitude $M(\delta_{ij})$ of a conformal correlation function of scalar fields via the definition\footnote{We remark that in momentum twistor language, this is a Mellin transform with respect to 4-brackets of the form $\ket{i\,i{+}1\,j\,j{+}1} \propto x_{ij}^2$.  However, from a twistor theorist's point of view, it might be interesting to contemplate instead a Mellin transform with respect to the variables $Z_i \cdot W_j = \ket{i\,j{-}1\,j\,j{+}1}$.  We thank D.~Skinner for this comment.}
	\be
	\left \langle
	\phi_{\Delta_1}(P_1) \ldots
	\phi_{\Delta_n}(P_n)\right \rangle
	=\,\oint \ud \delta_{ij}\  M(\delta_{ij})\, \prod_{i<j} \Gamma(\delta_{ij})\, P_{ij}^{-\delta_{ij}}. \label{Mellin}
	\ee
Let us look at this formula in detail. The main object in the above is $M(\delta_{ij})$, the Mellin amplitude. We conventionally defined it such that an overall product of $\Gamma$ functions always appears, and this will be convenient later. The Mellin amplitude is a function of the complex parameters $\delta_{ij}$ which are being integrated over a suitable\footnote{In all examples we consider, the integration contours may be taken parallel to the imaginary axis, with real parts chosen so that $\Re(\delta_{ij})>0$ for all $i,j$.} contour in the complex plane. These parameters are symmetric in their indices, and satisfy the constraints
	\be
	\delta_{ii}=-\Delta_i, \qquad \sum_{j} \delta_{ij}=0. \label{constraints}
	\ee
For more precise details on the measure we refer the reader to appendix~\ref{SymanzikSec}. Overall there are $\frac{n(n-3)}2$ independent parameters, and this is precisely the same as the number of independent cross-ratios.  This is not an accident, as the constraints~\reef{constraints} are precisely those of conformal invariance. This can be seen for instance by performing an inversion on the $x_i^\mu$ vectors, or more simply, by demanding that under $P_i \to \alpha_i P_i$ the overall amplitude scales like $\alpha_i^{\Delta_i}$. Upon solving the constraints, the Mellin representation becomes simply a product of familiar one-dimensional Mellin transforms, one for each cross-ratio. In particular, contour prescriptions are exactly the same as those for the one-dimensional transform, and the inverse Mellin transform is simply the product of the one-dimensional inverses\footnote{For sufficiently high number of legs there are ambiguities in the Mellin amplitude due to the vanishing of Gram determinants. In practice there is no cause for alarm, but see the discussion in~\cite{Fitzpatrick:2011hu}.}.

A nice way of thinking about these $\delta_{ij}$ parameters is as internal products of momenta~\cite{Mack:2009mi}. Indeed, if we parameterize
the $\delta_{ij}$ as
	\be
	\delta_{ij}=k_i\cdot k_j
	\ee
then the constraints~\reef{constraints} are automatically satisfied if $k_i^2=-\Delta_i$ and $\sum_i k_i=0$. In this way, one can think of a Mellin amplitude as depending on Mandelstam-like variables built out of these momenta, each momentum being associated with a given field. The resemblance of the $\delta_{ij}$ to momenta is not an accident and we will comment on it shortly.

What have we gained by using the representation~\reef{Mellin}? As it turns out we have gained considerably. Firstly, in contrast to the complicated functional dependence of general correlators in position space, Mellin amplitudes are simple meromorphic functions of their arguments $\delta_{ij}$. Secondly, the poles are directly related to the conformal block decomposition of the correlator. Consider for instance the four-point function. Applying the OPE decomposition in the $(12)$ channel, Mack~\cite{Mack:2009mi} has proven that the full Mellin amplitude ({\em i.e.} including gamma functions) has poles in the Mandelstam variable $s\equiv -(k_1+k_2)^2$,
	\be
	M(s,t)\simeq \sum_{p,n} \frac{P_n (t)\, C_{12p,n}\, C_{34p,n}}{s-(\Delta_p-l_p)-2n}. \label{decomposition}
	\ee
The position of the poles corresponds to the twist of the primary operators (labeled by $p$) appearing in the conformal block decomposition of the correlator. The extra $n$ summation in the above correspond to contributions from descendants of these fields. Further, the residues of these poles give products of the three-point couplings of the theory $C_{12p,n}$, up to a known polynomial in the remaining independent Mandelstam invariant $t=(k_1+k_3)^2$, a polynomial of order $l_p$, the spin of the primary field $p$. In this way, the Mellin amplitude makes the physical content of a given correlation function manifest: by examining its poles and their respective residues we can immediately determine which primary operators are involved and what their three-point couplings are.

The Mellin amplitudes defined above are valid for any scalar conformal correlation function. In practice, most of the work on this topic has so far focused on correlators computed with the help of the $AdS$/CFT correspondence. There, passage into Mellin space has allowed for a complete solution of the computation of tree-level correlation functions in arbitrary scalar field theories~\cite{Penedones:2010ue,Fitzpatrick:2011ia,Paulos:2011ie}. One finds that there are Feynman rules for directly evaluating the Mellin amplitude given a Witten diagram. These Feynman rules look remarkably similar to momentum space Feynman rules, once we set $\delta_{ij}=k_i\cdot k_j$. In this note we will focus on the computation of conformal integrals in flat space, but we will find the same structure at work. In particular, there seems to be a set of Feynman rules for the Mellin amplitudes of these conformal integrals, and the rules are considerably simpler than those found for $AdS$/CFT.

The resemblance of Mellin amplitudes to momentum space scattering amplitudes has been understood in the context of $AdS$/CFT. By considering the large energy limit of scattering in $AdS$, it is possible to show that the Mellin amplitude becomes precisely the $d+1$-dimensional flat-space scattering amplitude~\cite{Penedones:2010ue,Fitzpatrick:2011ia}. This means that we expect that there should be Feynman rules for computing the Mellin amplitude directly, and indeed this is the case. The rules were written in a compact form in~\cite{Paulos:2011ie} and shown to obey a BCFW-type recursion relation in~\cite{Fitzpatrick:2011ia}. These rules originally apply to the computation of the Mellin amplitude corresponding to tree-level Witten diagrams, but it is perhaps not too far-fetched to expect that similar rules should hold for conformal flat space integrals, and indeed we will find this is so.

\section{Conformal integrals in Mellin space}

\subsection{The polygon}

\label{sec:polygon}

We begin our labors with the calculation of the polygon integral in momentum space---a star in the dual position space. This integral is given by
	\be
	I_n\equiv \pi^{-h}\,\int \ud^d Q \prod_{i=1}^n \frac{\Gamma(\Delta_i)}{(-P_i \cdot Q)^{\Delta_i}}.
	\ee
The integration should be understood as $\int\ud^d Q\equiv \int_{-\infty}^{+\infty}\ud^d x$ (with $Q=\sqrt{2}(1,x^2,x^\mu)$ as explained in section~\ref{ambientspace}). We have also defined the convenient shorthand
	\be
	h\equiv \frac{d}2.
	\ee
The standard way of performing such integrals is to introduce Schwinger parameters, one per denominator factor,
	\be
	I_n=\pi^{-h}
	\int_0^{+\infty} \frac{\ud t_i}{t_i} t_i^{\Delta_i}
	\int \ud Q\,
	\exp\left(Q\cdot \sum_i t_i P_i\right).
	\ee
The $Q$ integral is gaussian, and we get
	\be
	I_n=\int_0^{+\infty} \prod_{i=1}^n \frac{\ud t_i}{t_i} t_i^{\Delta_i}
	\left(\sum_i t_i\right)^{-h}
	\exp\left(\frac{(\sum t_i P_i)^2}{2\,\sum t_i}\right).
	\ee
Now the point is that if the original integral satisfies the conformality condition $\sum_{i=1}^n \Delta_i = d$, we can drop factors of $\sum t_i$. A nice way to see this
is to introduce a partition of unity
	\be
	1=\int_{0}^{+\infty} \ud v\, \delta(v-\sum_i t_i).
	\ee
After a rescaling of the $t_i$ we get
	\be
	I_n=	\int_0^{+\infty} \frac{\ud v}v\, v^{\sum \Delta_i-h}
	\int_0^{+\infty}\prod_i \frac{\ud t_i}{t_i} t_i^{\Delta_i}
	\delta(1-\sum_i t_i) \exp\left(v \sum_{i<j} t_i t_j P_{ij}\right).
	\ee
If we now did the $v$ integral we would recover the usual Feynman parameter form of the loop integral, with the $t_i$ playing the role of Feynman parameters. Instead, we will do another rescaling, $t_i\to t_i/\sqrt{v}$ and perform the $v$ integral to get
	\be
	I_n=2
	\,
	\int_0^{+\infty}\prod_i \frac{\ud t_i}{t_i} t_i^{\Delta_i}
	\left(\sum_i t_i\right)^{\sum_{i=1}^n \Delta_i-d}
	\exp\left(\sum_{i<j} t_i t_j P_{ij}\right).
	\ee
We see that if the conformality condition is satisfied there is a simplification, and our integral becomes
	\be
	I_n=2\,\int_0^{+\infty}\prod_i \frac{\ud t_i}{t_i} t_i^{\Delta_i}
	\exp\left(\sum_{i<j} t_i t_j P_{ij}\right).
	\ee
To proceed we use Symanzik's trick~\cite{Symanzik:1972wj}, which tells us that%
	\be
	\int_0^{+\infty}\prod_i \frac{\ud t_i}{t_i} t_i^{\Delta_i} \exp(-\sum_{i<j} t_i t_j P_{ij})=\frac{1}2\oint \ud \delta_{ij} \prod_{i<j} \Gamma(\delta_{ij}) P_{ij}^{-\delta_{ij}} \label{symanzik}
	\ee
with $\delta_{ii}=-\Delta_i, \sum_{i}\delta_{ij}=0$, as explained in appendix~\ref{SymanzikSec}. These are the same $\delta_{ij}$ that we have introduced in section~\ref{mellinsec}. Recall that these constraints are solved by defining $\delta_{ij}=k_i \cdot k_j$, with $k_i^2=-\Delta_i$. Going back to our integral, we finally conclude that
	\be
	I_n=\oint \ud \delta_{ij} \prod_{i<j} \Gamma(\delta_{ij})  P_{ij}^{-\delta_{ij}}. \label{polymellin}
	\ee	
Comparing this expression with the definition of the Mellin amplitude~\reef{Mellin}, we come to the happy conclusion that the Mellin transform in the case is simply {\em one}, $M(\delta_{ij})=1$.

The reader might feel somewhat cheated by this result, but despite the simplicity of the Mellin amplitude the integral~\reef{polymellin} is definitely non-trivial; after all, there are still the factors of $\Gamma(\delta_{ij})$ left, which by convention we have left out of the definition of $M(\delta_{ij})$. However, this convention is useful since all conformal integrals we shall compute will always include the very same product of gamma functions. Their presence is related to the existence of ``double trace'' operators in the conformal block decomposition of the correlator, which always appear in such weak coupling computations\footnote{For a discussion of these issues in $AdS$ see for instance~\cite{ElShowk:2011ag}.}. This means knowing that the Mellin amplitude is one for all such conformal integrals is useful: it tells us how to straightforwardly write down their Mellin-Barnes representation without any further thought. In particular, the 4$d$ box integral~\reef{eq:1loopmassive} is simply
\begin{equation}
{\hbox{\lower 12pt\hbox{
\begin{picture}(30,30)
\put(0,0){\includegraphics[width=30pt]{1loopmassive.pdf}}
\end{picture}
}}}=
P_{13} P_{24} \,\oint \ud \delta_{ij}\, \prod_{i<j}^4 \Gamma(\delta_{ij})\, P_{ij}^{-\delta_{ij}},
\end{equation}
and similarly the fully massive $n$-gon integral in $n$ dimensions is a similarly simple $\frac{1}{2}n(n-3)$-dimensional Mellin integral.
In order to connect with previously known results from the literature we consider in detail the cases
$n=4$ and $n=6$ in appendices~\ref{box} and~\ref{hex} respectively.

There is another more interesting reason why we have chosen the convention~\reef{Mellin} for defining the Mellin amplitude. This is because the fact that the Mellin amplitude as we've defined it is simply a constant (in this case) is analogous to the fact that in momentum space, non-derivative contact interactions also correspond to trivial amplitudes---they are constants as well (up to momentum-conserving delta functions). The natural question that arises is whether this analogy will continue to hold when we have a conformal integral corresponding to an exchange diagram. An example of such an integral is the scalar double box integral in 4 dimensions, which is dual to a position space $3\to 3$ exchange diagram in $\phi^4$ theory, shown in Fig.~\ref{fig:2loopmassive}(b). It is precisely to this kind of integral that we shall consider next, and indeed we shall find that this analogy continues to hold perfectly.

\subsection{Two polygons with one common edge}

Let us consider then a two-loop computation. Once again we will keep our calculations generic, and consider not a double box but rather an $n$-gon and an $m$-gon, as before, glued together along one edge. The integral in the dual position space is now
	\be
	I_{n,m}=\frac{\pi^{-2h}}2\,\int \ud Q_1 \ud Q_2\, \prod_{i\in L}^n \frac{\Gamma(\Delta_i)}{ (- P_i\cdot Q_1)^{\Delta_i}}\,
	\prod_{j\in R}^m\frac{\Gamma(\Delta_j)}{ (- P_j\cdot Q_1)^{\Delta_j}} \frac{1}{(- Q_1\cdot Q_2)^\delta},
	\ee
where $L,R$ refer to the left and right polygons. The only restriction is that both integrals are conformal, so that $\sum_{i\in L} \Delta_i+\delta=\sum_{i \in R} \Delta_i+\delta=d$. As before we introduce Schwinger parameters and compute the $Q$ integrals. Conformality simplifies matters and we end up with
%
%
	\be
	I_{n,m}=\frac{2}{\Gamma(\delta)}
	\int_0^{+\infty}\prod_i \frac{\ud t_i}{t_i} t_i^{\Delta_i}
	\int_0^{+\infty} \frac{\ud s}{s} s^\delta \,
	e^{(1+s^2)P_L^2+2 s P_L\cdot P_R+ P_R^2}
	\ee
with $P_{L,R}=\sum_{i\in L,R}t_i P_i$. Exactly as in the previous case, we use Symanzik's formula~\reef{symanzik} to directly derive the Mellin representation from the above integral. But this time, because of the ``internal'' Schwinger parameter $s$, which came from exponentiating the $(-Q_1\cdot Q_2)^{-\delta}$ factor, we get a non-trivial Mellin amplitude
	\bea
	M(\delta_{ij})&=& \frac{1}{\Gamma(\delta)}\int_0^{+\infty} \frac{\ud s}s s^{\delta-\sum_{i\in L, j\in R} \delta_{ij}} (1+s^2)^{-\sum_{i<j \in L} \delta_{ij}}\nonumber \\
	&=&\frac{1}{\Gamma(\delta)}\int_0^{+\infty} \frac{\ud s}s s^{\delta-s_{L}}(1+s^2)^{-\frac 12 (\sum_{i\in L} \Delta_i-s_L)}	
	\eea
with $s_L\equiv -(\sum_{i \in L} k_i)^2$. Performing the integral we find the simple result
	\be
\label{above}
	M(s_L)=
	\frac{
	\Gamma\left(\frac {\delta-s_L}2\right)
	\Gamma\left(\frac {d-2\delta}2\right)}
	{2\,\Gamma(\delta)\, \Gamma\left(\frac{d-\delta-s_L}2\right)}.
	\ee
To better exhibit the pole structure of the Mellin amplitude, we can write it instead as
	\be
	M(s_L)=-\,\sum_{n=0}^{+\infty}\, \frac{\left(1+\delta-h\right)_n}{n! \Gamma(\delta)}\, \frac{1}{s_L-\delta-2\,n}.\label{meldoublebox}
	\ee
This result is in precise accord with the general expectations of section~\ref{mellinsec}. As predicted by Mack, we find poles corresponding to the primary field of dimension $\delta$ being exchanged, plus an infinite series of poles labeled by $n$ corresponding to its descendants. The Mellin amplitude therefore looks like precisely a momentum space scattering amplitude, in this case an amplitude involving exchange of an infinite set of massive fields, with particular propagator normalization factors.

We should not forget however, that at the end of the day our interest is in analyzing the conformal integrals appearing in SYM theory scattering amplitudes. In particular, the fully massive double box of Fig.~\ref{fig:2loopmassive} corresponds to setting $\delta=1$ and $d=4$. When we do this, the infinite sum beautifully collapses onto the single term
	\be
\label{blah1}
	M(s_L)=\frac{1}{1-s_L}
	\ee
Why is this so? Notice that the first argument of the Pochhammer symbol $(1+\delta-h)_n$ vanishes if $\delta=h-1=d/2-1$: in other words if the conformal dimension $\delta$ is that of a free field. But for free fields we have $\Box \phi=0$, which effectively kills off the contributions from descendant fields to the expression~\reef{above}.

To summarize, we have found that the fully massive two-loop double box integral, which appears to be too difficult to evaluate
in position space with current methods, can be computed
by substituting
\be
\label{eq:recalling}
s_L=-(k_1+k_2+k_3)^2=3-2(\delta_{12}+\delta_{13}+\delta_{23})
\ee
into~\reef{blah1} and~\reef{Mellin}, yielding
the extremely simple representation
\begin{equation}
\label{eq:doubleboxmellin}
{\hbox{\lower 16pt\hbox{
\begin{picture}(50,37)
\put(0,0){\includegraphics[width=50pt]{2loopmassive.pdf}}
\end{picture}
}}}=
\frac{1}{2} P_{14} P_{25} P_{36}
\,\oint \ud \delta_{ij}\  \frac{1}{ \delta_{12} +  \delta_{13} +  \delta_{23}-1} \, \prod_{i<j} \Gamma(\delta_{ij})\, P_{ij}^{-\delta_{ij}}
\end{equation}
as a Mellin integral of dimensionality $\frac{1}{2} 6 (6-3) = 9$.
In section~\ref{feynmanrules} we argue that this result generalizes to higher loops in a straightforward way to
provide a $(2L-1)(L+1)$-dimensional Mellin representation of the fully massive $L$-loop ladder integral.

\subsection{Massless limits of Mellin amplitudes}
\label{sec:massless}

We have emphasized that for Mellin amplitudes it seems most natural to always begin with the most massive version of any
integral of interest.  This means that all $x_i$ are taken to be arbitrary, which from the point of view of conformal correlation
functions is the most natural thing to do.  However many of the integrals appearing in scattering amplitudes require special
constraints on the kinematics---often one or more pairs of cyclically adjacent $x$'s are lightlike separated from each other
($x_{i,i+1}^2 = 0$), or even exactly equal to each other ($x_i = x_{i+1}$).
The question naturally arises then as to how one can specialize to such cases since the definition of the Mellin
amplitude~\reef{Mellin} seems to break down if any $x_{ij}^2 = 0$.

In many cases (including all of the examples we consider explicitly in the following sections), taking the limit as some $x_{ij}^2 \to 0$ can be done quite easily at the level of the Mellin amplitude.  This happens whenever the associated $\ud \delta_{ij}$ contour integral reduces, in the limit, to the contribution from the single pole at $\delta_{ij} = 0$.
In such cases the prescription for setting some $x_{ij}^2 \to 0$ is therefore simply:  (a) omit the factor $\Gamma(\delta_{ij}) P_{ij}^{-\delta_{ij}}$ from the product, (b) drop the integration over $\delta_{ij}$, and (c) set $\delta_{ij} = 0$ everywhere else it appears in the integrand.

Unfortunately things are not always so simple, notably in the presence of denominator factors such as in the result~\reef{eq:doubleboxmellin}.  In that example we begin with a 9-fold integral for a quantity depending on 9 independent cross-ratios.  Suppose we then want to take the limit described in the caption of Fig.~\ref{fig:2loopmassive} to recover the 4-mass double box, whose value is given explicitly by the formula~\reef{eq:lloopmassive} in terms of the 2 remaining cross-ratios which survive in this limit.
What happened to the other 7 cross-ratios?  It is easy to check that 2 of them go to zero in the limit.  These two are simple to deal
with using the rule explained in the previous paragraph.  However, 5 of the original 9 cross-ratios go to 1 in the limit, leaving us with a 7-fold representation for a quantity depending on only 2 cross-ratios.

Let us conclude these remarks by noting that in examples where this kind of thing happens, typically it is possible to explicitly evaluate some of the `extra' Mellin integrals via successive applications of Barnes lemmas. For instance in the double box case one easily reduce down to four Mellin integrals--the same number as the representation used in~\cite{Usyukina:1992jd}. Once we can no longer continue in this fashion, a possibility would be to trade Mellin integrals for Euler integrals ({\em i.e.} some integrals from $0$ to $1$) by using identity~\reef{meltoeuler}, since these might be then easier to perform explicitly.

\section{Feynman rules for conformal integrals}
\label{feynmansec}
\subsection{Feynman rules in Mellin space}
\label{feynmanrules}

In the last two sections we have seen that the box and double box loop integrals have Mellin amplitudes which are extremely simple: they are respectively $1$ and $1/(1-s)$, where $-s$ is the square of the total Mellin momentum flowing through the internal propagator in the dual exchange graph ({\em i.e.} Fig~\ref{fig:2loopmassive}(b)). This is in agreement with Mack's analysis of the Mellin amplitude, as presented in section~\ref{mellinsec}. These two simple results suggest that there might exist some simple ``Feynman rules'' for writing down Mellin amplitudes. As we shortly reviewed in section in~\ref{mellinsec}, such rules were indeed found to exist for calculations performed in $AdS$/CFT. The rules were written in a compact form in~\cite{Paulos:2011ie} and, for Mellin amplitudes of tree-level scalar correlators in $AdS$, they take the form:
	\begin{itemize}
	\item For each internal line of conformal dimension $\Delta$ in the diagram write down a propagator
		\be
		\frac{1}{n! \Gamma(1+\Delta+n-h)}\, \frac{1}{k^2+\Delta+2n}.
		\ee
	\item In $g^{(m)} \phi^m$ theory\footnote{Here we really mean interactions without derivatives between arbitrary scalars {\em e.g.} $\phi_1 \phi_2 \phi_3 \phi_4$, where each field can have a different conformal dimension.} the vertex connecting lines with dimensions $\Delta_i$ is given by
\begin{multline}
	V_{n_i}^{\Delta_i}= g^{(m)}\,\Gamma\left(\frac{\sum_{i=1}^m \Delta_i-2h}2\right) \prod_{i=1}^m \left(1+\Delta_i-h\right)_{n_i}\nonumber\\
	 \times F_A^{(m)}\left(\frac{\sum_{i=1}^n \Delta_i\!-\!2h}{2},\{-\!n_1,\!\ldots \!,-\! n_m\},\{1\!+\!\Delta_1\!-\!h,\!\ldots\!,1\!+\!\Delta_n\!-\! h\},1,\ldots,1\right),
\end{multline}
with $F_A^{(m)}$ one of Lauricella's multivariable hypergeometric functions.
	\item The Mellin amplitude is obtained by attributing momenta to every line consistent with momentum conservation at every vertex and summing over all $n_i$.
	\end{itemize}
Now consider taking the limit where the conformality condition $\sum_{i=1}^m \Delta_i=d$ is satisfied, keeping dimensions otherwise arbitrary. In this limit, the Lauricella function simplifies to one, simplifying substantially the vertex factors. However, the overall gamma function is divergent as well. If one goes through the computation in detail, one sees that these gamma functions originally arise from the integration over the radial coordinate of $AdS$. When the conformality condition is satisfied, the integral becomes divergent due to the near horizon region. This indicates that in this limit, indeed the $AdS$ integral is reducing to a boundary one. To deal with this divergence one can simply absorb the divergence into the definition of the coupling constant $g^{(m)}$.

Since the Lauricella function has simplified, the vertex factors take a nice factorized form, where every line connected to it contributes a Pochhammer symbol. Therefore we may as well associate these factors to the propagators instead of the vertices. The modified rules become:
	\bea
	\mbox{Vertex}&:& \qquad \hat g^{(m)} \\
	\mbox{Propagator}&:&\qquad \frac{\left(1+\Delta-h\right)_n}{n! \Gamma\left(1+\Delta-h\right)}\, \frac{1}{k^2+\Delta+2\,n}
	\eea
We see that this is almost exactly the same as in~\reef{meldoublebox}. The difference can be explained by different normalization conventions for propagators, and is easily ammended. With this final modification, we are led to the strong suspicion that the conformal flat space integrals have Mellin amplitudes described by Feynman-like rules, where to each internal line one associates a factor as in~\reef{meldoublebox}.

Our results have direct consequences for the computation of loop integrals with dual conformal symmetry. They imply that for diagrams which look like tree-level position space correlators in $\phi^4$ theory, one can immediately write down the corresponding Mellin amplitude. To reiterate the rules are:
	\begin{itemize}
	\item To every external line in the dual diagram, attribute a Mellin momentum $k_i$, satisfying $k_i^2=-1$.
	\item Mellin momentum is conserved at each vertex.
	\item To every internal line attribute a propagator factor:
		\be
		\frac{1}{p^2+1}
		\ee
	where $p$ is the total Mellin momentum flowing through that line.
	\end{itemize}
We should emphasize that these rules only hold for {\em tree-level} graphs in the dual position space. In the original momentum space variable, window-like diagrams would not be captured by the above set of rules. We will have more to say about this in the discussion section. Another important point is that these rules are really only appropriate for conformal integrals without nontrivial numerator factors (such as the one shown in Fig.~\ref{fig:magicnumerator}), which so far we have not considered. Of course many such integrals are important for the computation of scattering amplitudes, and this does not mean that the Mellin space approach is useless there. Rather as we will see in later sections, for cases with nontrivial numerators we will end up with a kind of mixed representation, where the index structure of the numerators ends up separate from the bulk of the integral, the latter being expressed in Mellin space.

This is a somewhat unaesthetic state of affairs. Although we will not fully solve it in this note, an attempt to remedy it is given in appendix~\ref{genmel}, where we define a generalized version of the Mellin transform. The basic idea is to think of numerators as fields with negative (integer) conformal dimension. With our definition, any one-loop conformal integral with an arbitrary number of numerators has a generalized Mellin transform which is simply equal to one. This is a promising start, and we hope to explore this further in future work. For now let us turn to the exploration of some of the interesting consequences of the existence of Feynman rules for the Mellin amplitude.
\subsection{Consequences of the Feynman representation}
\label{consequences}
We have found that the Mellin amplitude that corresponds to a tree-level diagram in position space is given by a product of factors. For instance for diagrams in $\phi^4$ theory the Mellin amplitude becomes a product of simple propagator-like factors. This is quite nice, since products of Mellin amplitudes translate into convolutions of functions in position space. A quick reminder of how this works in a one-dimensional example: suppose we have two functions $f(x), g(x)$ with Mellin transforms $M^f(s), M^g(s)$. That is,
	\be
	M^f(s)=\int_0^{+\infty} \frac{\ud x}x\, x^{s}\, f(x), \qquad M^g(s)=\int_0^{+\infty} \frac{\ud x}x\, x^{s}\, g(x).
	\ee
Then the function $h(x)$ which corresponds to the product of the two Mellin amplitudes is given by
	\bea
	h(x)&=&\oint \frac{\ud s}{2\pi i}\, M^f(s) M^g(s) x^{-s}=\oint \frac{\ud s}{2\pi i}\,
	\int_0^{+\infty} \frac{\ud y}y\, y^{s}\, f(y)M^g(s) x^{-s}\nonumber \\
	&=&\int_0^{+\infty} \frac{\ud y}y\, f(y) g(x/y). \label{piece1}
	\eea
This means that we can break up the calculation of the position space functions into convolutions of simpler functions. The integral representations one obtains in this fashion can be thought of as solutions to certain differential equations that the original conformal integral satisfies.

For clarity, consider for instance the double box integral of Fig.~\ref{fig:2loopmassive}(b). We have found that this integral can be represented in Mellin form by
	\be
	I_{3,3}=\oint \ud \delta_{ij}\, \frac{1}{1-s}\, \prod_{i<j}\Gamma(\delta_{ij})\, P_{ij}^{-\delta_{ij}}, \label{piece2}
	\ee
up to a prefactor irrelevant to the current discussion. Inspection of the diagram~\ref{fig:2loopmassive}(b) shows that by acting with the laplacian operator on the internal line we should reduce the integral to something resembling a six-point star, or hexagon integral in the original momentum variables. In Mellin space it is clear how to see this, since polygon integrals correspond to setting the Mellin amplitude to one. Therefore we must come up with a differential operator which cancels the propagator factor $1-s$ in the above. Start by defining the homogeneous derivative
	\be
	\dij=P_{ij}\frac{\partial}{\partial P_{ij}}.
	\ee
Inside the Mellin integration sign this derivative gives $-\delta_{ij}$. Recalling~\reef{eq:recalling}
we can then write
	\be
	2\left(1+\sum_{i<j}^3 \dij\right)\bigg[I_{3,3}\bigg]=I_6=\oint \ud \delta_{ij}\, \prod_{i<j}\Gamma(\delta_{ij})\, P_{ij}^{-\delta_{ij}}. \label{difeqi33}
	\ee
The conformally invariant functions
	\bea
	\hat I_{3, 3}(u_i)&\equiv& (P_{14}\, P_{25}\, P_{36})\, I_{3,3},\nonumber \\
	\hat I_{6}(u_i)&\equiv &(P_{14}\, P_{25}\, P_{36}) I_{6}\label{confi}
	\eea
depend on the same cross-ratios $u_i$,
	\bea
	u_i&=&u_{i,i+3}, \qquad\quad i=1,2,3, \label{ui61}\\
	u_{i+3}&=& u_{i+1,i+5}, \qquad i=1,\ldots, 6 \label{ui62}
	\eea
with the notation of equation~\reef{eq:uij},
	\be
	u_{ij}\equiv \frac{P_{i,j+1}\, P_{i+1,j}}{P_{ij}\,P_{i+1,j+1}}.
	\ee
In terms of these conformally invariant functions we can write~\reef{difeqi33} as
	\be
	u_3\, \partial_{u_3}\, \hat I_{3,3}=\frac{1}{2}\, \hat I_6 \label{difeq}.
	\ee
Such a differential equation implies that knowing $I_6$ should be sufficient to solve $I_{3,3}$. The way to do this is precisely by the convolution method presented above. We shall need the position space expression corresponding to the Mellin amplitude $1/(1-s)$:
	\be
	\oint \ud \delta_{ij}\, \frac{1}{1-s}\, \prod_{i<j}^6\, (P_{ij})^{-\delta_{ij}}=-\frac{1}{P_{14}\,P_{25}\, P_{36}}\, \frac{\theta(1-u_3)}{2}\prod_{i\neq 3} \delta(1-u_i) \label{piece3}
	\ee
with $\theta(x)$ the step function.
Putting together~\reef{piece1}, \reef{piece2} and~\reef{piece3} we can write
	\be
	\hat I_{3, 3}(u_i)=-\frac{1}{2} \int_{u_3}^{+\infty} \frac{\ud u_3'}{u_3'}\, \hat I_6(u_1,u_2,u_3',\ldots,u_9)
	\ee
which indeed solves~\reef{difeq}.

It is clear that this kind of strategy will generalize to the case of the $L$-loop ladder diagram, which will have $L-1$ propagator
factors in its Mellin amplitudes and consequently will be able to be expressed as an $(L-1)$-fold integral of the scalar $2L+2$-gon integral. It should also be clear that our arguments are more general, and hold not only for scalar box integrals, but for more complicated integrals with arbitrary legs and conformal dimensions---only the convolving function becomes slightly more complicated.

\section{The chiral pentagon}
\label{pentasec}
\subsection{The pentagon in Mellin space}

Here we consider the chiral pentagon integral shown in Fig.~\ref{fig:penthex}(a), which is a basic ingredient in all one-loop MHV
amplitudes in SYM theory~\cite{ArkaniHamed:2010gh}.
We start with the related integral

	\be
	I_n^{1}=\frac{1}{\pi^2}\,\int \ud^4 Q\, (-Q\cdot Y)\prod_{i=1}^n\,\frac{1}{\left(- P_i \cdot Q\right )} \label{onenum}
	\ee
(the superscript indicates the presence of 1 numerator factor).
Up to a certain normalization factor to be fixed shortly, the chiral pentagon is clearly a particular case of the above---it corresponds to imposing the requirements mentioned in section~\ref{sasec},
	\be
\label{eq:vanishing}
	Y\cdot P_i=0, \qquad i=1,\ldots, 4,
	\ee
and further demanding that $P_{12}=P_{34}=0$ in order to match the kinematics of Fig.~\ref{fig:penthex}(a).

To perform the integral~\reef{onenum}, we introduce Schwinger parameters as usual,
	\be
	I_n^{1}=\pi^{-h}\,
	\int \prod_{i=1}^n \frac{\ud t_i}{t_i} t_i^{\Delta_i}
	\int \ud^4 Q\, (-Q \cdot Y)\, e^{Q\cdot\sum t_i P_i}.
	\ee
The trick now is to trade the unexponentiated $Q$ for a derivative with respect to the argument of the exponential. The calculation then proceeds in a straightforward fashion, and in fact it is essentially the same as for the one-loop integral without numerator which we carried out in section~\ref{sec:polygon}.

With the conditions~\reef{eq:vanishing}, the Mellin representation of the integral is then simply
	\bea
\label{eq:pentagon2}
	I_5^{1} &=& (-Y\cdot P_5) \oint \ud \delta_{ij} \prod_{i<j} \Gamma(\delta_{ij}) P_{ij}^{-\delta_{ij}}\label{pentaconf},
	\eea
with
\begin{equation}
\label{pentagonconstraints}
\sum_{j \neq i} \delta_{ij} = \begin{cases}1 & i \ne 5,\\ 2 & i=5\end{cases}.
\end{equation}
So, from the above we read off the Mellin amplitude
which once again is equal to one (or, more precisely, $(-Y \cdot P_5) \times 1$).

The result~\reef{eq:pentagon2} is valid for general chiral pentagons. A generic five-point conformal integral depends on five arbitrary cross-ratios. We choose these as
	\be
	u_i\equiv \frac{P_{i,i+3}\, P_{i+1,i+2}}{P_{i,i+2}\,P_{i+1,i+3}}, \qquad i=1,\ldots,5
	\ee
where cyclicity is understood. If we now specialize to the kinematics of Fig.~\ref{fig:penthex}(a) then the only non-zero cross-ratios are $u_1,u_3$ and $u_4$.
In order to compare with~\reef{eq:pentagonresult} it remains only to fix the overall normalization of~\reef{onenum} compared
to~\reef{eq:pentagonlabeling}, which is uniquely fixed by normalizing the leading singularity to 1.
This requires a little work and the nontrivial identity
\begin{equation}
\ket{12(345) \cap (678)} \ket{1247} = \ket{1245} \ket{1267} \ket{3478} \left[
1 - u_3 - u_4 + u_1 u_3 u_4\right]
\end{equation}
(where the momentum twistors appearing inside the brackets are labeled according to~(\ref{eq:pentagonlabeling}))
which may be derived for example with the help of appendix~A of~\cite{Prygarin:2011gd}.

This leads to our final expression
\begin{equation}
\label{eq:pentagon3}
{\hbox{\lower 16pt\hbox{
\begin{picture}(50,50)
\put(0,0){\includegraphics[width=50pt]{pentagon.pdf}}
\end{picture}
}}}=
P_{14} P_{25} P_{35} (1-u_3 -u_4+u_1u_3u_4)
\,\oint \ud \delta_{ij}\  \prod_{i<j} \Gamma(\delta_{ij})\, P_{ij}^{-\delta_{ij}}.
\end{equation}
This expression provides a simple representation of the chiral pentagon integral as a three-dimensional Mellin integral (from the $\frac{1}{2} 5 (5-3) = 5$
original $\delta_{ij}$ we must set $\delta_{12}=\delta_{34} = 0$, leaving only three integration variables---of course the $i,j = 1,2$ and $i,j=3,4$ terms are omitted from the product).
It is now trivial to verify numerically that this expression agrees precisely with~(\ref{eq:pentagonresult}), and in fact we derive it analytically in appendix~\ref{pentappendix}. However we will see in the next section that in fact the special numerator factor leads to an easier way to do the calculation.

\subsection{Mellin magic numerators}

Numerators of the type shown in Fig.~\ref{fig:magicnumerator} were called `magic numerators' in~\cite{Drummond:2010mb}.  In this section we will see, by careful study of the factor $1-u_3-u_4+u_1 u_3 u_4$ appearing in the example~\reef{eq:pentagon3}, that these numerators can perform certain magic also in Mellin space.  Specifically we will see that this factor seems carefully constructed to trivialize one of the contour integrals.
After starting with the three-dimensional integral~(\ref{eq:pentagon3})  and performing one of the Mellin integrals we obtain
	\bea
{\hbox{\lower 16pt\hbox{
\begin{picture}(50,50)
\put(0,0){\includegraphics[width=50pt]{pentagon.pdf}}
\end{picture}
}}}
\,&= &(1-u_3-u_4+u_1 u_3 u_4) \oint \frac{\ud c_3\, \ud c_4}{(2\pi i)^2} u_3^{-c_3} u_4^{-c_4}\,M(c_3,c_4), \nonumber \\
	M(c_3,c_4)&\equiv&\gn{1-c_3}^2 \gn{c_3}^2  \gn{1-c_4}^2\gn{c_4}^2\ _2 F_1(c_3,c_4,1,1-u_1). \label{ipent}
	\eea
The $c_3$ and $c_4$ integrals run along the imaginary axis with a small positive real part. In this way, for $|u_3|$ and $|u_4|$ smaller than one we can close the contour on the left and pick up the poles at $c_3=-n_1$ and $c_4=-n_2$, for all $n_1,n_2$
non-negative integers.

Now take the prefactor and place it inside the contour integral. Next perform changes of variables in $c_3$ and $c_4$ such that all terms come multiplied by $u_3^{-c_3}$ and $u_4^{-c_4}$. Of course, once we do this, the contours are not necessarily the same. However, assume for a moment they would be. Then we are free to add up the different terms under the same integral sign, and we get exactly zero! This means that the original integral should be equal to the terms we have lost by the shifting of the contour.  This is more clearly seen with a simple example. Consider the integral
	\be
	(1-x)\, \int_{\epsilon-i\infty}^{\epsilon+i \infty}\frac{\ud s}{2\pi i}\, \gn{s}^2 \gn{1-s}^2\, x^{-s}
	\ee
with $0<\epsilon<1$. For $0<|x|<1$, we close the contour on the left and pick up the poles at $s=-n$, with $n$ a non-negative integer. The result is found to be $-\log(x)$. Now let us do the calculation in a different way. By placing the prefactor inside the integral and doing a change of variables we find
	\bea
	&&(1-x)\, \int_{\epsilon-i\infty}^{\epsilon+i \infty} \frac{\ud s}{2\pi i}\, \gn{s}^2 \gn{1-s}^2\, x^{-s}\nonumber \\
	&&\qquad=\int_{\epsilon-i\infty}^{\epsilon+i \infty} \frac{\ud s}{2\pi i}\, \gn{s}^2 \gn{1-s}^2\, x^{-s}+\int_{\epsilon-1-i\infty}^{\epsilon-1+i \infty} \frac{\ud s}{2\pi i}\, \gn{1+s}^2 \gn{-s}^2\, x^{-s} \nonumber \\
	&&\qquad= 0+\left(\int_{\epsilon-1-i\infty}^{\epsilon-1+i \infty}-\int_{\epsilon-i\infty}^{\epsilon+i \infty}\right)\left( \frac{\ud s}{2\pi i}\, \gn{s}^2 \gn{1-s}^2\, x^{-s}\right)\nonumber \\&&\qquad =-\log(x).
	\eea
In the last step we used the fact that the contour in the last integral only encloses the pole at $s=0$. In other words, in this case the prefactor is so special that the result of the integral becomes the residue of a single pole, as opposed to an infinite sum.

This is precisely what is going on in our more complicated example of the pentagon integral~\reef{eq:pentagon3}. There we have four terms coming from the prefactor $1-u_3-u_4+u_1 u_3 u_4$. Multiplication by the last three of these four terms involves shifting the contour, and just as in our simple example we get the zero mentioned previously plus the terms coming from the contour shift. The result is
	\bea
	{\hbox{\lower 16pt\hbox{
\begin{picture}(50,50)
\put(0,0){\includegraphics[width=50pt]{pentagon.pdf}}
\end{picture}
}}}
\,&=&\left\{\mbox{Res}_{c_3=0}\,\oint \frac{\ud c_4}{2\pi i} u_3^{-c_3} u_4^{-c_4}  M(c_3+1,c_4)- \right.\nonumber \\
	&& \left.-u_1\, \mbox{Res}_{c_4=0}\,\oint \frac{\ud c_3}{2\pi i} u_3^{-c_3} u_4^{-c_4}  M(c_3+1,c_4+1)+ (c_3\leftrightarrow c_4)\right \}+ \\
	&& + u_1\, \mbox{Res}_{c_3=0,c_4=0}\, u_3^{-c_3} u_4^{-c_4}  M(c_3+1,c_4+1). \label{resi5}
	\eea
Evaluating the residues leads to
	\bea
	&& \tilde I_5=\log(u_1)\log(u_3){+}\log(u_3)\log(u_4){+}\log(u_4)\log(u_1)
	+\, u_1 \frac{\partial^2}{\partial \alpha\, \partial \beta}\bigg[\ _2 F_1(\alpha,\beta,1,1{-}u_1)\bigg]_{\alpha=\beta=1} \nonumber \\
	&& -\oint \frac{\ud c}{2\pi i}\, \left(u_3^{-c}{+}u_4^{-c}\right)\, \gn{1{-}c}^2\, \gn{c}^2\, \frac{\ud}{\ud \alpha}\bigg[\ _2 F_1(c,\alpha,1,1{-}u_1){-}u_1\, _2 F_1(1{+}c,\alpha,1,1{-}u_1)\bigg]_{\alpha=1}.
	\eea
To proceed we need to evaluate the derivatives of the hypergeometric function with respect to its parameters. Such derivatives can be easily evaluated by first using the series representation for the hypergeometric function, taking the derivative, and performing the sum. In this way we find
	\bea
	\frac{\ud}{\ud \alpha}\bigg[\ _2 F_1(c,\alpha,1,1-u_1)-u_1\, _2 F_1(1+c,\alpha,1,1-u_1)\bigg]_{\alpha=1} =\frac{1-u_1^{-c}}c,\\
	\bigg[\ _2 F_1(\alpha,\beta,1,1-u_1)\bigg]_{\alpha=\beta=1}
	=\sum_{n=0}^{+\infty} (1-u_1)^n H_n^2=\frac{\log(u_1)^2+\Li_2(1-u_1)}{u_1}
	\eea
where $H_n\equiv \sum_{k=1}^n\, k^{-1}$ is the $n$-th harmonic number. We get
\begin{multline}
	 \tilde I_5=\oint \frac{\ud c}{2\pi i} (u_1^{-c}-1)(u_3^{-c}+u_4^{-c})\, \gn{1-c}\gn{c}^2\gn{-c}\\
+\log(u_1)\log(u_3)+\log(u_3)\log(u_4)+\log(u_4)\log(u_1)+\log(u_1)^2+\Li_2(1-u_1). \label{representation}
\end{multline}
The single Mellin integral can be easily performed and we recover the well-known result for $\tilde I_5$ shown in~\reef{eq:pentagonresult}.

To conclude this section, we recall that in~\cite{Drummond:2010cz} it was found that the pentagon integral satisfies the particularly simple differential equation
	\be
\label{eq:pentagondiffeq}
	u_3 u_4 \partial_{u_3} \partial_{u_4} I_5^{(1)}=1.
	\ee
Using the representation~\reef{representation} it is trivial to see this. Clearly this operator annihilates the single Mellin integral, since in it $u_3$ and $u_4$ appeared summed, not multiplied. The action of the operator on the remaining terms immediately gives $1$. We note that to obtain this result it is actually not necessary to compute the derivatives of hypergeometric functions. One can start directly from expression~\reef{ipent}, apply the differential operator, and shift contours. One gets an expression analogous to~\reef{resi5}, and quite trivially one immediately obtains one as the answer.

\section{The chiral hexagon and differential operators}
\label{hexsec}
Now we consider the hexagon in Fig.~\ref{fig:penthex}(b). In the ambient space formalism this integral translates into
(again, up to an overall normalization factor which will be fixed below)
	\be
	I_6^{2}=\frac{1}{2\pi^2}\,\int \ud^4 Q \,\frac{(- Q\cdot Y)(- Q\cdot Y')} {\prod_{i=1}^6 (- P_i \cdot Q)} \label{chiralhex}
	\ee
where the special vectors $Y,Y'$ satisfy
	\bea
\label{eq:specialdef}
	Y\cdot P_i&=&0,\qquad i=1,\ldots,4,\\
	Y'\cdot P_i&=&0, \qquad i=3,\ldots,6,
	\eea
and further we demand $P_{i,i+1}=0$ for all $i$. There are two possible solutions to each of the above equations, for a total of four different integrals.  However two of these are related to the other two, leaving only two independent choices.  Our focus here will be on the
chiral integral shown in in Fig~\ref{fig:penthex}(b) and~\reef{eq:hexagon} in which both numerators are of the same type.

As always we begin with a more general $n$-gon integral, this time with two numerators,
	\be
	I_n^{2}=\frac{Y^A\, Y^B}{\pi^d}\int \ud^4 Q\, Q_A\, Q_B\, \prod_{i=1}^n\frac{ \Gamma(\Delta_i)}{(-P_i \cdot Q)^{-\Delta_i}},\label{twonums}
	\ee
and again we assume the conformality condition $\sum_i^n \Delta_i=d+2$, but keep all vectors arbitrary. To do this integral we follow the same strategy as for the pentagon, obtaining:
	\bea
\label{eq:hexagon1}
	I_n^{2}&=& Y_A Y'_B \oint \ud \delta_{ij} \, T^{AB}\prod_{i<j} \Gamma(\delta_{ij}) P_{ij}^{-\delta_{ij}}\label{in2}
	\eea
with
	\bea
\label{eq:hexagon2}
	T^{AB}&\equiv & \eta^{AB}+\sum_{i\neq j} \delta_{ij} \frac{P_i^A P_j^B}{P_{ij}}+\sum _{i} P_i^A P_i^B \hat S_{i}.
	\eea
Here we have defined $\hat S_{i}$ as the operator which modifies the constraints on the $\delta_{ij}$ in such a way that $\Delta_i\to \Delta_i+2$, keeping all other dimensions fixed. This somewhat technical point will not be relevant for the chiral hexagon. Notice that $T_{AB}$ is symmetric and traceless as it should be, due to the fact that
	\be
	\eta^{AB}T_{AB}=d+2-\sum_{i\neq j}\delta_{ij}=d+2-\sum_{i=1}^n \Delta_i=0.
	\ee
There is a very nice way of rewriting equation~\reef{in2}. To see this notice that inside the Mellin integral, we have $\delta_{ij}=-P_{ij}\frac{\partial}{\partial P_{ij}}\equiv - \hat \partial_{ij}$. Therefore we can write
	\be
\label{eq:hexagon7}
	I_n^{2}=\, Y_A\, Y'_B\left(\eta^{AB}-\sum_{i,j} \frac{P_i^A P_j^B}{P_{ij}} \hat \partial_{ij}+\sum _{i} P_i^A P_i^B \hat S_{i}\right)\left[ \oint \ud \delta_{ij} \prod_{i<j} \Gamma(\delta_{ij})\, P_{ij}^{-\delta_{ij}}\right].
	\ee
That is, the two-numerator polygon conformal integral can be written as a certain differential operator acting on the expression between square brackets. The object being acted upon looks exactly like the polygon integral {\em without} numerators, namely~\reef{polymellin}. The only catch is that for the polygon integral we have $\sum_i^n \Delta_i=d$, whereas here, on account of the original two numerators, we must have $\sum_i^n \Delta_i=d+2$ for conformality. Therefore the object appearing above could never arise from the computation of a $d$-dimensional conformal integral.
Instead it must come from a $d+2$-dimensional integral dependent on the same collection of cross-ratios (which are largely ignorant of
the dimensionality of the space they live in)\footnote{They may satisfy polynomial relations known as Gram determinant constraints, but these have no effect on any of our analysis.}.  This is the Mellin space manifestation of the well-known connection between scalar integrals
in $d+2$ dimensions and tensor integrals in $d$ dimensions (see for example~\cite{Bern:1993kr}).

Let us now go back to the particular case of the chiral hexagon in $d=4$. The result can be written as a particular operator acting on the hexagon integral in $d=6$. The generic integral depends on nine independent conformal cross-ratios, as introduced in equations~\reef{ui61} and~\reef{ui62}. In the kinematic regime relevant to the chiral hexagon we have $P_{i,i+1}=0$ for all $i$, and so the cross-ratios $u_4$ through $u_9$ actually vanish. The Mellin representation of the integral~\reef{eq:hexagon}
including all overall factors, and with the numerator factor appearing in~\reef{eq:hexagon1} and~\reef{eq:hexagon2}
written out
explicitly in terms of momentum twistors is given by
\begin{equation}
\label{eq:hexagon4}
{\hbox{\lower 35pt\hbox{
\begin{picture}(80,80)
\put(0,10){\includegraphics[width=60pt]{hexagon.pdf}}
\put(12,3){\makebox(0,0){$1$}}
\put(46,3){\makebox(0,0){$6$}}
\put(12,76){\makebox(0,0){$3$}}
\put(46,76){\makebox(0,0){$4$}}
\put(-8,40){\makebox(0,0){$2$}}
\put(67,40){\makebox(0,0){$5$}}
\end{picture}
}}}=
P_{15} P_{24} P_{36} \,
\oint \ud \delta_{ij}\  M(\delta_{ij}) \prod_{i<j} \Gamma(\delta_{ij})\, P_{ij}^{-\delta_{ij}}
\end{equation}
where (see appendix~\ref{numeratorappendix} for details)
\begin{equation}
\label{eq:hexagonnumerator}
M(\delta_{ij}) =
1 - \frac{\ket{1345} \ket{2346} }{ \ket{1346} \ket{2345}} \delta_{15}
- \frac{\ket{1356}\ket{2346}}{\ket{1346}\ket{2356}} \delta_{25}
- \frac{\ket{1246}\ket{1345}}{\ket{1245}\ket{1346}} \delta_{14}
- \frac{ \ket{1246} \ket{1356} }{ \ket{1256} \ket{1346}} \delta_{24}
\end{equation}
and of the $\frac{1}{2} 6(6-3) = 9$, the six of the form $\delta_{i,i+1}$ are set to zero (and the corresponding
terms omitted from the product), leaving only
three independent integration variables.
We have checked numerically that the expression~\reef{eq:hexagon4}
agrees precisely with~\reef{eq:hexagonresult}.

Alternatively, we can also derive the differential operator form of the chiral hexagon. Since the integral~\reef{chiralhex} is conformally invariant, we can trade derivatives with respect to $P_{ij}$ by derivatives with respect to cross-ratios, and in the end we obtain
\begin{multline}
{\hbox{\lower 35pt\hbox{
\begin{picture}(80,80)
\put(0,10){\includegraphics[width=60pt]{hexagon.pdf}}
\put(12,3){\makebox(0,0){$1$}}
\put(46,3){\makebox(0,0){$6$}}
\put(12,76){\makebox(0,0){$3$}}
\put(46,76){\makebox(0,0){$4$}}
\put(-8,40){\makebox(0,0){$2$}}
\put(67,40){\makebox(0,0){$5$}}
\end{picture}
}}}=\bigg(u_2(1{-}u_1{-}u_3){+}u_2(1{-}u_1) u_1\hat \partial_{u_1} \\
 {+}u_2(1{-}u_3)\,u_3 \partial_{u_3}{-}(1{-}u_2)(1{-}u_1{-}u_3)\,u_2\partial_{u_2}\bigg)\, \mathcal I_6(u_1,u_2,u_3) \label{difeqn}
\end{multline}
where the zero-mass $d=6$ hexagon
	\be
	\mathcal I_6(u_1,u_2,u_3)\equiv \left(P_{14}\, P_{25}\, P_{36}\, I_6\right)_{u_4=\ldots=u_9=0}
	\ee
has been given explicitly\footnote{Regrettably, the notations of~\cite{DelDuca:2011jm} and~\cite{ArkaniHamed:2010gh} are inconsistent. We stick with the latter, so Del Duca et al's $u_1, u_2, u_3$ are our $u_2, u_3, u_1$ respectively.} in~\cite{DelDuca:2011jm} (but see also appendix~\ref{hex}).
Of course, this particular differential equation relating the chiral two-numerator hexagon in $4d$ to the massless scalar hexagon in $6d$ is very well-known~\cite{Drummond:2010cz} (and see also~\cite{Dixon:2011ng} for further applications).

\section{Conclusion and discussion}

Motivated by the success of Mellin representations~\cite{Mack:2009mi} for
studying correlation functions in general CFTs
and in particular in $AdS$/CFT, we have here initiated a preliminary
investigation of the suitability of using Mellin
representations for dual conformal integrals of the type appearing
in SYM theory scattering amplitudes\footnote{As opposed to
more general Mellin-Barnes representations, which are a general tool
for a much wider class of Feynman integrals.}.
We have explored this possibility by working out explicit, and very
simple, Mellin amplitudes for several particular
integrals.  In those cases for which results are available in the
literature,
including the four-mass box, the four-mass double box,
and
the chiral pentagon and hexagons, we have checked agreement (either
analytically or numerically)
between these results and our Mellin representations.  Beyond this,
we have seen how simple it is to write Mellin representations
for large classes of integrals which seem far beyond the ability to
evaluate in $u$-space with currently available methods.
Examples of this include the fully massive $L$-loop ladder diagram,
or even the fully massive
one-loop $n$-gon integral in $n$ dimensions\footnote{For even $n$, an
explicit formula for the symbol of this integral is given
in~\cite{Spradlin:2011wp}, however the fully massive
version of the integral has not yet been evaluated for any $n>4$.},
whose Mellin representation, when normalized according to the
standard convention~\reef{Mellin}, is just 1!

\begin{figure}
\begin{center}
\includegraphics[width=100pt]{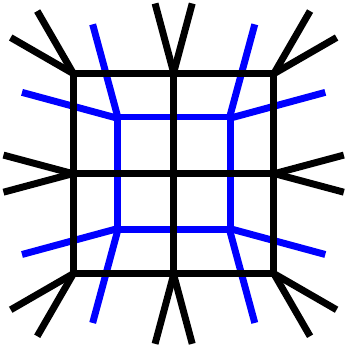}
\end{center}
\label{fig:window}
\caption{The simplest example of a dual conformal `window' diagram (black) whose
dual diagram (blue) has a loop in $x$-space.
Here we have shown the fully massive version of the integral;
the fully massless version contributes to the four-loop four-particle
MHV amplitude in SYM theory.}
\end{figure}

This was made possible by introducing a set of Feynman rules for
conformal integrals, and showing how they can be interpreted as
arising from the Feynman rules
found in~\cite{Paulos:2011ie,Fitzpatrick:2011ia} for $AdS$ correlation
functions. Unfortunately, this analogy with momentum space amplitudes
does not hold at the quantum level, which is to say whenever there are
loops in $x$-space (window-like diagrams in the original momentum space,
see Fig.~\ref{fig:window}). This is because as we saw in
section~\ref{mellinsec}, the Mellin amplitude can have only simple
poles, whereas momentum space loops have branch cuts. Of course this
doesn't mean that there aren't some simple rules for writing down
Mellin amplitudes involving $x$-space loops---indeed it is extremely
likely those must exist---it is just that they have yet to be worked
out in detail. For some work along this direction in the context of
$AdS$ computations see~\cite{Penedones:2010ue,Fitzpatrick:2011hu}.

It would be particularly interesting to work out if there exist similar
diagrammatic rules for integrals involving general numerators of the
type reviewed in section~\ref{sasec}. In the examples we have studied we
have seen that for integrals with only a single numerator, the numerator
of the Mellin representation is just a constant prefactor (independent
of the integration variables $\delta_{ij}$), while for integrals with
more than one numerator factor a non-trivial numerator can appear such
as in~\reef{eq:hexagon4}.  Because these numerator factors depend
explicitly on cross-ratios, these Mellin amplitudes are not literally
the same thing one would obtain if one computed the inverse Mellin
transform of the integral's $u$-space answer.  Rather they are some
kind of hybrid representation (akin to writing a function of $x$ as a
Fourier transform of something which depends on $k$ but also explicitly on
$x$).   This is not necessarily a bad thing; if for example there are
simple and physically well-motivated general rules for writing such
hybrid representations, then we are all for it.  Alternatively, the
generalized Mellin transform introduced
in appendix~\ref{genmel} might also be a natural object to consider.

A particularly important role is played by the one-loop scalar $n$-dimensional $n$-gon integral, whose Mellin amplitude
is exactly $1$ (in the normalization which is now standard in the $AdS$/CFT literature). We have seen in the concrete example of the chiral hexagon
how the numerator factors transform into certain differential operators acting on these simpler integrals. Generically, the statement is that a $d$-dimensional conformal integral containing an even number $2n$ of numerators can always be rewritten as an operator acting on the same conformal integral without numerators in a higher dimension. This operator contains up to $n$ derivatives with respect to the $P_{ij}$, and in turn these will transform into derivatives with respect to cross-ratios. In this way Mellin space allows for an easy deduction of differential relations
amongst multi-loop integrals.

Going in the other direction, we have shown that certain higher-loop integrals with trivial numerator factors may be written
as integral operators acting on the $n$-gon.  Our chief example is a formula for the 6-mass two-loop
double box integral as a very simple integral of the 6-mass scalar hexagon in six dimensions,
\begin{equation}
{\hbox{\lower 17pt\hbox{
\begin{picture}(50,37)
\put(0,0){\includegraphics[width=60pt]{2loopmassive.pdf}}
\end{picture}
}}}(u,\ldots)=
-\frac{1}{2} \int_u^{+\infty} \frac{du'}{u'}
{\hbox{\lower 27pt\hbox{
\begin{picture}(60,60)
\put(0,0){\includegraphics[width=60pt]{massivehexagon.pdf}}
\put(30,30){\makebox(0,0){$d{=}6$}}
\end{picture}
}}}(u',\ldots)
\end{equation}
where $u = x_{13}^2 x_{46}^2/(x_{14}^2 x_{36}^2)$, found in
section~\ref{consequences}.
Here the diagrams on the left- and right-hand sides are each functions
of 9 cross-ratios, eight of which (represented
by $\ldots$) are identified on both sides.  It is only the appearance of
one preferred $u$ which breaks the cyclic
symmetry of the hexagon integral.  Again we emphasize that explicit
formulas in $u$-space are currently not known for either diagram,
but the double box integral on the left does appear in two-loop SYM
theory scattering amplitudes.
This formula clearly generalizes to one which expresses the fully massive $L$-loop ladder
diagram as an $L-1$-fold integral of the fully massive scalar $2L+2$-gon.

Another natural arena in which to explore these methods is in the context of the chiral double pentagon integral, which was evaluated
in~\cite{Dixon:2011nj} as an ingredient in
the analytic formula for the two-loop 6-point NMHV amplitude presented in that paper.
It satisfies known differential equations relating it
to both the chiral hexagon and the six-dimensional scalar hexagon~\cite{Drummond:2010cz,Dixon:2011ng}, which manifest themselves
very simply in Mellin space.
Based on our analysis we expect that it should also be possible to express the double pentagon
as a first-order differential operator acting on a single integral of the 8-dimensional octagon.  
This relation is especially interesting in light of the special role played by a set of chiral octagon integrals~\cite{ArkaniHamed:2010gh} which
provide a basis for one-loop integrands and have been evaluated analytically.
These chiral octagons should themselves be expressible as certain differential operators acting on the 8-dimensional octagon.  This suggests
a possible link between the double pentagon integral and these chiral octagons, which would surely be worth further investigagion.

To summarize, it is clearly important to better understand the structure of the scalar $n$-gon integral in $n$ dimensions, whose Mellin amplitude is just 1.  Since all numerator factors in a general integral can be written as differential operators acting on this object, and all denominator factors can be written as integral operators acting on this object, we believe that all (at least, all non-window) dual conformal integrals relevant to SYM theory can be written as integro-differential operators acting on the scalar $n$-dimensional $n$-gon.

On a different note, it would be very interesting to understand, for those
integrals which can be expressed
in terms of generalized polylogarithm functions, if there is a simple
way to read off the symbol of an integral directly
from its Mellin representation (or even to be able to look at a Mellin
representation and quickly determine what degree of transcendentality,
if any, the resulting integral has).  Presumably, if there is a way to determine this it will be by careful examination of the arguments of the gamma functions which appear generically in any Mellin transform. Indeed, (poly)logarithms can only arise from sums of double or higher order poles in the Mellin amplitude.  A first step would surely be to understand the $n$-dimensional scalar $n$-gon integrals, which have Mellin transform $1$ but a very nontrivial symbol~\cite{Spradlin:2011wp}.

Keeping our wildest speculation for the very end, it is perhaps our
greatest hope that
it might be possible to find a recursion relation
which works directly at the level of
Mellin representations for multi-loop amplitudes in SYM theory, akin to
the BCFW-type recursion relation
which is known to hold at the level of the
integrand~\cite{CaronHuot:2010zt,ArkaniHamed:2010kv,Boels:2010nw}.
If this hope was
realized, and one could generate Mellin representations for arbitrary
amplitudes `at the touch of a button', then our suggestion
that Mellin space might serve as a useful stepping stone between
integrands and integrals would be fully realized.

\appendix

\section{The Symanzik star formula}
\label{SymanzikSec}

For completeness, in this section we review the Symanzik star integration formula in Euclidean space as discussed in ~\cite{Mack:2009mi}. For a proof and more details we refer the reader to the original reference~\cite{Symanzik:1972wj}. Consider a set of $n$ points in Euclidean space $x_i$ and their differences $x_i-x_j$. In the embedding formalism we have $P_{ij}\equiv -P_i\cdot P_j=(x_i-x_j)^2$. Symanzik's formula is then
	\bea
	 \int_{0}^{+\infty}\!\! \left(\prod_{i=1}^n \frac{\ud t_i}{t_i} t_i^{\Delta_i}\right) e^{-\left(\sum_{1\leq i<j \leq n}\!\! t_i t_j\, P_{ij}\right)}&=&
	\frac{1}2\,\oint \ud \delta_{ij}\! \prod_{1\leq i<j \leq n}\!\!\Gamma(\delta_{ij})\,P_{ij}^{-\delta_{ij}}.
	\eea
This identity can be deduced by using the Cahen-Mellin integral $e^{-x}=\int \frac{\ud s}{2\pi i} \,\Gamma(s)\, x^{-s}$ for each factor $\exp(t_i t_j Q_{ij})$ and performing the Schwinger parameter integrals. The integration measure on the right-hand side is more precisely given by
	\be
	\oint \ud \delta_{ij}=2\,\int_{-i \infty}^{+i\infty}\prod_{1\leq i<j\leq n}  \ud \delta_{ij}\, \prod_{i=1}^n \delta(\Delta_i-\sum_{j\neq i} \delta_{ij}).
	\ee
We have included the factor of two since in the majority of cases one solves for a subset of the $\delta_{ij}$ as independent variables, and in that case it is easy to see that a factor of $1/2$ arises from the Jacobian coming from the delta functions. Of course, more general choices are possible. To see this, first notice that the Dirac delta functions imply constraints on the $\delta_{ij}$:
\be
\sum_{i\neq j} \delta_{ij}=\Delta_j \equiv -\delta_{ii}\label{dij}
\ee
for all $i$.  Now pick a particular solution of the set of equations~\reef{dij}, $\delta^0_{ij}$.  Then we write
	\be
	\delta_{ij}=\delta^0_{ij}+\sum_{k=1}^{\frac 12 n(n-3)}\,c_{ij,k} s_k
	\ee
with
	\be
	c_{ii,k}=0,\qquad \sum_{j\neq i} c_{ij,k}=0.
	\ee
Choosing as independent variables the $\left (\frac 12 n(n-3)\right)^2$ coefficients $c_{ij,k}$ with $2\leq i < j \leq n$ (with the exception of $c_{23,k}$), with the further restriction $|\mbox{det} \, c_{ij,k}|=1$, we can write
	\be
	\int \ud \delta_{ij}\to 2\,\int_{-i \infty}^{+i\infty} \prod_{k=1}^{\frac 12 n(n-3)} \frac{\ud s_k}{2\pi i}.
	\ee
The integration paths are chosen parallel to the imaginary axis, with real parts such that the real parts of the arguments of the gamma functions are positive.

\section{Useful hypergeometric function identities}
\label{identities}
In the appendices we show how one can (sometimes)
easily go from the Mellin representation to position space. We will be using the following basic identities
	\bea
	\ _2 F_1(a,b,c,z)&=&\frac{\Gamma(c)}{\Gamma(b) \Gamma(c-b)}
	\int_0^1 \frac{\ud t}{t(1-t)}\, t^b (1-t)^{c-b} (1-t z)^{-a},\label{hyper1} \\
	\ _2 F_1(a,b,c,z)&=&
	\oint \frac{\ud s}{2\pi i} \,\frac{\Gamma(c) \Gamma(s)\Gamma(c-a-b+s)\Gamma(a-s)\Gamma(b-s)}{\Gamma(a)\Gamma(b)\Gamma(c-a) \Gamma(c-b)}(1-z)^{-s},\label{hyper2} \\
	\ _2 F_1(a,b,c,z)&=&\frac{\Gamma(c)}{\Gamma(a)\Gamma(b)}
	\oint \frac{\ud s}{2\pi i}\, \frac{\Gamma(s)\Gamma(a-s)\Gamma(b-s)}{\Gamma(c-s)}\, (-z)^{-s}.\label{hyper3}
	\eea
{}From the first two equations above we also get the useful relation
\bea
	&& \oint \frac{\ud s}{2\pi i}\, \Gamma(a_1+s)\Gamma(a_2+s)\Gamma(b_1-s)\Gamma(b_2-s)\, z^{-s} \nonumber \\
	&&=\Gamma(a_1+b_1)\Gamma(a_2+b_2)
	\int_0^1 \frac{\ud t}{t(1-t)}\, t^{b_2+a_1}(1-t)^{a_2+b_1}\, [1-t(1-z)]^{-b_1-a_1}. \label{meltoeuler}	
\eea
which allows to go from a Mellin-type integral to an Euler-type one.
\section{Selected details}
\subsection{The general box integral}
\label{box}
The box integral is written in Mellin form as
	\be
	I_{4}\equiv \oint \ud \delta_{ij} \, \prod_{i<j}^4\, \Gamma(\delta_{ij}) P_{ij}^{-\delta_{ij}}.
	\ee
Defining the cross-ratios
	\be
	u\equiv \frac{ P_{12}\, P_{34}}{P_{13}\, P_{24}} \qquad
	v\equiv \frac{P_{14}\, P_{23}}{P_{13}\, P_{24}}
	\ee
and the quantities
	\be
	\Delta^+_{ij}\equiv \frac{\Delta_i+\Delta_j}2, \qquad \Delta^{-}_{ij}\equiv \frac{\Delta_i-\Delta_j}2,
	\ee
the integral becomes
	\bea
	I_4&=&\frac{1}{(-2\, P_{12})^{\Delta_{12}^+}
	\, (-2\, P_{34})^{\Delta_{34}^+}}
	\,
	\left(
	\frac{P_{24}}{P_{14}}
	\right)^{\Delta_{12}^-}
	\,
	\left(
	\frac{P_{14}}{P_{13}}
	\right)^{\Delta_{34}^-}	\, f_4(u,v)
	\eea
with
\begin{multline}
	f_4(u,v)\equiv \oint \frac{\ud s \, \ud t}{(2\pi i)^2}
	\Gamma(\Delta_{12}^+ +s)\Gamma(\Delta_{34}^+ + s)
	\Gamma(-\Delta_{12}^- -s-t)\Gamma(\Delta_{34}^{-}-s-t)\nonumber \\
	\times	\Gamma(t)\Gamma(\Delta_{12}^- - \Delta_{34}^- +t)\, u^{-s}\, v^{-t}.
\end{multline}
Using the integral representations for the hypergeometric function it is straightforward to obtain
\begin{multline}
	 f_4(u,v)=\frac{\Gamma(\s{12}{-}\df{12})\,
	\Gamma(\s{12}{-}\df{34})\, \Gamma(\s{34}{+}\df{12})\, \Gamma(\s{34}{+}\df{34})}{\Gamma(\s{12}{+}\s{34})}
	\int_0^1 \frac{\ud t}{t(1{-}t)}\, (1{-}t)^{\s{34}{-}\df{12}}\, \nonumber \\
	 \times \, t^{\s{12}{+}\df{34}}\, (1{-}t(1{-}u))^{-\s{12}{+}\df{12}}
	\,
	\ _2 F_1\left(\s{12}{-}\df{12},\s{34}{+}\df{34},\s{12}{+}\s{34},
	1{-}\frac{(1{-}t)\, t\, v}{1{-}t(1{-}u)}\right). \nonumber
\end{multline}
The above integral can be done in terms of Appell $F_4$ functions, but we will not do so here. Instead, we specialize to the case of equal conformal dimensions, $\Delta_i=n$. Then we obtain
	\bea
	f_4(u,v)&\to &\frac{\Gamma(n)^4}{\Gamma(2\,n)}\int_0^1 \frac{\ud t}{t(1-t)}\,\left(\frac{t\, (1-t)}{1-t(1-u_1)}\right)^n\,
	_2 F_1\left(n,n,2n,
	1- \frac{(1-t)\, t\, v}{1-t(1-u)}\right)\nonumber \\
	&=& \frac{\Gamma(n)^4}{\Gamma(2\,n)}\int_0^1 \frac{\ud t}{t(1-t)}\, v^{-n}\ _2 F_1\left(n,n,2n,1-
	\frac{1-t(1-u)}{t(1-t) v}\right).
	\eea
The case relevant for $d=4$ is $n=1$, whereupon the integral reduces to
	\be
	\int_0^1 \ud t\, \frac{\log\left(\frac{1-t(1-u)}{v t(1-t)}\right)}{1-t(1-u)-v\, t(1-t)}.
	\ee
It is now a straightforward matter to integrate this expression to obtain the final result~(\ref{eq:lloopmassive}).

\subsection{The six-dimensional one-mass hexagon}
\label{hex}
In this section we derive a simple Mellin-Barnes representation for the hexagon integral in 6 dimensions. The Mellin amplitude is one, and so the integral which we denote $I_6$ is given by
	\be
	I_6=\oint \ud \delta_{ij}\, \prod_{i<j}\Gamma(\delta_{ij}) P_{ij}^{-\delta_{ij}}
	\ee
By conformality, up to some prefactor, the six-point integral can depend only on $6\times 3/2=9$ cross-ratios $u_i$. In the notation of~\reef{eq:uij} we take these to be given by
	\bea
	u_i&=&u_{i,i+3}, \quad \qquad i=1,2,3 \\
	u_{i+3}&=& u_{i+1,i+5}, \qquad i=1,\ldots, 6.
	\eea
Accordingly the constraints on $\delta_{ij}$ can be solved for nine independent variables which we denote by $c_i$. There are many possible choices for which variables to choose, and this corresponds to shifting the function of cross-ratios by some monomial in the cross-ratios. One particular choice leads to
	\bea
	&&\hat I_6\equiv (P_{14}P_{25}P_{36})\, I_6=\oint \prod_{i=1}^9 \frac{\ud c_i}{2\pi i}\, u_i^{-c_i}\,  \left(\prod_{i=4}^9 \Gamma(c_i)\right)\, \Gamma(c_1-c_5-c_6)\Gamma(c_2-c_6-c_7)\nonumber \\
	&& \times \Gamma(1-c_1-c_3+c_5+c_8)\,
	\Gamma(c_3-c_7-c_8)\, \Gamma(c_1-c_8-c_9)\, \Gamma(1-c_1-c_2+c_6+c_9)\nonumber \\
	&& \times \Gamma(1-c_2-c_3+c_4+c_7)\, \Gamma(c_3-c_4-c_5)\, \Gamma(c_2-c_4-c_9),
	\eea
To proceed we will focus on a specific kinematic regime, corresponding to the one-mass case. In this limit all cross-ratios except $u_1,u_2,u_3$ and $u_4$ vanish. The corresponding Mellin integrals become trivial, as one simply picks up one residue for each of the cross-ratios. The integral becomes
\begin{multline}
	\hat I_6=-\oint\left(\prod_{i=1}^4\,\frac{\ud c_i}{2\pi i}\, u_i^{-c_i}\right)\, \Gamma(c_1)^2\,\Gamma(c_2)\Gamma(c_3)\Gamma(c_4) \, \Gamma(1-c_1-c_2)\nonumber \\
\times	\Gamma(1-c_1-c_3)\,\Gamma(c_2-c_4)\,
	\Gamma(c_3-c_4)\Gamma(1-c_2-c_3+c_4).
\end{multline}
By making good use of the identities of section~\ref{identities} we can simplify this to
\begin{multline}
	\hat I_6=-\oint \frac{\ud c_2\, \ud c_3}{(2\pi i)^2}\,u_2^{-c_2}\, u_3^{-c_3}\, \frac{\Gamma(1-c_2)^3\,\Gamma(c_2)^2\, \Gamma(1-c_3)^3 \Gamma(c_3)^2}{\Gamma(2-c_2-c_3)}\,\nonumber \\
	 \times _2 F_1\left(1\!-\!c_2,1\!-\!c_3,2\!-\!c_2\!-\!c_3,1\!-\!u_1\right)\ 
	_2 F_1\left(c_2,c_3,1,1-u_4\right).
\end{multline}
This provides an efficient numerical representation for the integral. Of course nothing stops us from trading the Mellin-Barnes integrals for Euler integrals. This can be done for instance by using the Euler representation of the two hypergeometric functions and performing the Mellin-Barnes integrals. This leads to the alternative representation
	\bea
	\hat I_6 &=&\int_0^1 \ud s\, \ud t\, \frac{\log\left(\frac{1-s(1-u_1)}{ (1-s)(1-t(1-u_4))\, u_2}\right)}{
	\bigg(s u_1+(1-s)[1-(1-t(1-u_4))u_2]\bigg)\bigg((1-t) s \,u_3+t\bigg)}.
	\eea
We have checked that this agrees numerically with the exact analytical result of reference~\cite{DelDuca:2011jm}, which was denoted there by $\mathcal I_{6,m}$ In doing the comparison note that our cross-ratios correspond to the ones in that reference after we change variables such that
\be
\hat I_6(u_1,u_2,u_3,u_4)=\mathcal I_{6,m}(u_2,u_3,u_1,u_4).
\ee
%

\subsection{The chiral pentagon}
\label{pentappendix}

There are five Mellin integrals in the expression for $I_5$. After solving the constraints on the $\delta_{ij}$ we find we can write
\begin{multline}
	I_5=\frac{(-Y\cdot P_5) }{ P_{25}\, P_{14}\,P_{35}}
	\oint \prod_{i=1}^5 \left(\frac{\ud c_i}{(2\pi i}  u_i^{c_i-1} \Gamma(1-c_i)\right) \gn{c_1+c_2-c_4-1}\nonumber \\
	 \times \gn{c_3+c_4-c_1}\gn{c_2+c_3-c_5}\gn{c_1+c_5-c_3-1}\gn{c_4+c_5-c_2}
\end{multline}
with the cross-ratios
	\be
	u_i=u_{i,i+2}=\frac{P_{i,i+3}\, P_{i+1,i+2}}{P_{i,i+2}\,P_{i+1,i+3}}, \qquad i=1,\ldots,5.
	\ee
We are interested in evaluating the Mellin integrals for the particular kinematics corresponding to $P_{12}=P_{34}=0$ (as in Fig.~\ref{fig:penthex}(a)), which implies $u_2=u_5=0$. In this limit we simply pick up  the residues at $c_2=c_5=1$, obtaining
\begin{multline}
	\oint \frac{\ud c_1 \ud c_3 \ud c_4}{(2\pi i)^3} u_1^{c_1-1}u_3^{c_3-1} u_4^{c_4-1} \gn{1-c_1}\gn{1-c_3}\gn{1-c_4}\nonumber \\
	\gn{c_1-c_3}\gn{c_3}\gn{c_1-c_4}\gn{c_4} \gn{c_3+c_4-c_1}.
\end{multline}
Now we perform the $c_1$ integral using the identity~\reef{hyper2}, to get
	\be
	\oint \frac{\ud c_3 \ud c_4}{(2\pi i)^2} u_3^{c_3-1} u_4^{c_4-1} \gn{1-c_3}^2 \gn{c_3}^2 \gn{1-c_4}^2\gn{c_4}^2\ _2 F_1(1-c_3,1-c_4,1,1-u_1).\nonumber
	\ee
By further using the representation~\reef{hyper1} for the hypergeometric function, the $c_3$ and $c_4$ integrals can also be performed, finally giving
	\bea
	&&\int_0^1 \ud t \frac{\log\left( u_3[1-t(1-u_1)]\right)}{\left(1-u_3[1-t(1-u_1)]\right)\left(u_4(1-t)+t\right)}= \nonumber
	\\
	&&\qquad \frac{1}{1-u_3-u_4+u_1 u_3 u_4}\left[ \log(u_1)\log\left(\frac{1-u_1}{1-u_1 u_4}\right)-\log(u_3) \log (u_4)-\Li_2(1-u_3)\right.\nonumber \\
	&&\qquad\left. +\Li_2(1-u_1 u_3)-\Li_2\left(\frac{1-u_4}{1-u_1 u_4}\right)+\Li_2\left(\frac{u_1(1-u_4)}{1-u_1 u_4}\right)\right].
	\eea
One can check numerically that the expression between brackets is exactly~\reef{eq:pentagonresult}, and that the prefactor in front of the expression in square brackets is exactly the appropriate relative normalization factor appearing as the overall factor in~(\ref{eq:pentagon3}).

\subsection{The Mellin numerator of the chiral hexagon}
\label{numeratorappendix}

Here we provide some details on the derivation of the Mellin numerator~\reef{eq:hexagonnumerator} for the chiral hexagon integral.
We begin by solving the constraints~\reef{eq:specialdef}, which in $x$-coordinates read
\bea
(y - x_i)^2 &=& 0, \qquad i=1,2,3,4, \\
(y'-x_i)^2 &=& 0, \qquad i=3,4,5,6.
\eea
Details on solving this kind of problem in general may be found in~\cite{ArkaniHamed:2010gh}.  There are two solutions
for each of $y$ and $y'$:
\bea
y &=& (13) \quad {\rm or} \quad (612) \cap (234), \\
y' &=& (46) \quad {\rm or} \quad (345) \cap (561),
\eea
where we express the 4-dimensional vectors as antisymmetric products of momentum twistors
as in~\reef{eq:mt},
labeling the legs and faces of the hexagon as shown in~\reef{eq:hexagon} and Fig.~\ref{fig:penthex}(b)
respectively.
Amongst the four choices of $\{y,y'\}$ two pairs are related by parity.  The chiral hexagon we are interested in corresponds
to choosing the same type of solution for each numerator, so we will proceed with
\begin{equation}
\label{eq:using}
y = (612) \cap (234),
\qquad
y' = (345) \cap (561).
\end{equation}

Now we turn our attention to the relevant portion of the prefactor in~\reef{eq:hexagon7}, which we can write as
\begin{equation}
(y-y')^2 \left[ 1 - \sum_{i,j} \frac{(y - x_i)^2 (y'-x_j)^2}{(y-y')^2 (x_i-x_j)^2} \delta_{ij}\right].
\end{equation}
Only four of the cross-ratios appearing in the sum are nonzero, and those are easily
computed using~\reef{eq:using}, which gives exactly~\reef{eq:hexagonnumerator} together with the overall factor
\begin{equation}
(y - y')^2 = \ket{1256} \ket{1346} \ket{2345} = (-P_{15}) (+P_{36}) (-P_{24})
\end{equation}
appearing in~\reef{eq:hexagon4}.
It may be of interest to note that
two of these cross-ratios can be written easily in terms of the standard $u_i$ defined in~\reef{ui61} and~\reef{eq:uij}:
\begin{equation}
\frac{\ket{1356}\ket{2346}}{\ket{1346}\ket{2356}} =1-u_3, \qquad
\frac{\ket{1246}\ket{1345}}{\ket{1245}\ket{1346}} = 1-u_1.
\end{equation}
The other two cross-ratios (the first and fourth in~\reef{eq:hexagonnumerator}) are
the two roots of the quadratic equation
\begin{equation}
u_2 x^2 - (1-u_1+u_2-u_3) x + (1 - u_1 - u_3 + u_1 u_3) = 0.
\end{equation}

\section{A generalized Mellin transform}
\label{genmel}

Suppose one wants to compute an integral with a general numerator structure of the form
	\be
I^{N,M}\equiv\, \frac{\prod_{i=1}^N \Gamma(\Delta_i)}{\pi^h \prod_{a=1}^M m_a!}\, \int \ud^d Q\, \frac{\prod_{a=1}^M (Q\cdot Y_a)^{m_a}}{\prod_{i=1}^N (Q\cdot P_i)^{\Delta_i}}.
	\ee
The trick is to introduce Schwinger parameters for the denominators and complex integrals for numerators,
	\be
	I^{N,M}=\frac{1}{\pi^h}\, \int \ud^d Q\,\int_0^{+\infty} \prod_{i=1}^N \frac{\ud t_i}{t_i} t_i^{\Delta_i}\, \oint \prod_{a=1}^M\, \frac{\ud z_a}{(2\pi i)}\, z_a^{-1-m_a}\, \exp[ Q\cdot(\sum_i t_i P_i+\sum_a z_a Y_a)].
	\ee
The integral in $Q$ is then simply performed. Conformality makes annoying factors of $\sum t_i+\sum z_j$ drop out. The net result is that we get
	\be
	I^{N,M}=2\,\int_0^{+\infty} \prod_{i=1}^N \frac{\ud t_i}{t_i} t_i^{\Delta_i}\, \oint \prod_{j=1}^M\, \frac{\ud z_j}{(2\pi i)}\, z_j^{-1-m_a}\,\exp[(\sum_i t_i P_i+\sum_j z_j Y_j)^2].
	\ee
Now we do Symanzik's trick, slightly generalized. We slice up the $P_{ij}$ exponentials and introduce Mellin integrations $\delta_{ij}$ for each of them, as in~\ref{SymanzikSec}. For the cross terms $t_i z_a P_i\cdot Y_a$ and for $z_a z_b Y_a Y_b$ we substitute the exponentials by their series representation, introducing sums over integers $q_{ia}$ and $n_{ab}$. Notice that the parameter $q_{ij}$ has indices running over different ranges.  The integration over the $t's$ gives Dirac deltas imposing constraints on $\delta_{ij}$ Mellin parameters; the complex $z_i$ integrations give Kronecker deltas imposing constraints on the $n_{ab}$ parameters and we get
	\be
	I^{N,M}=\, \sum_{n_{ab},q_{ia}} \oint \ud \delta_{ij}\, \prod_{i<j}^N \Gamma(\delta_{ij})\,\left(P_i\cdot P_j\right)^{-\delta_{ij}}\,
	\prod_{a<b}^M \frac{1}{n_{ab}!}\, \left(Y_a\cdot Y_b\right )^{n_{ab}}
	\,
	\prod_{i,a} \frac{1}{q_{ia}!}\,\left(P_i\cdot Y_{a}\right)^{q_{ia}}.
	\ee
The sums and the integral satisfy the constraints
	\bea
	\sum_{j\neq i} \delta_{ij}-\sum_{a} q_{ia}=\Delta_i,\\
	\sum_{b\neq a} n_{ab}+\sum_{i} q_{ia}=m_a.
	\eea
This example suggests it is natural to introduce a generalized Mellin transform for any conformal integral, by simply adding a general function $M(\delta_{ij},q_{ia},n_{ab})$ in the integrand above. The generalized Mellin amplitude of any of the conformal integrals $I^{N,M}$ is one. With similar calculations it is not too hard to show that the Mellin amplitude of an exchange integral is a simple pole, but the location of the pole depends on the quantities $n_{ab}$ and $q_{ia}$. In other words, everything works out as if we could attribute negative conformal dimensions to numerator factors.

\acknowledgments

AV thanks D.~Nandan and C.~Wen for discussions, and
MS and AV are especially grateful to S.~Raju for several helpful and stimulating conversations. MP acknowledges useful discussions with S. El-Showk, J.~Kaplan, G.~Korchemski, and P.~Vanhove.
This work was supported in part by the LPTHE, Universit\'e Pierre et Marie Curie (MP); the US Department of Energy under contracts DE-FG02-91ER40688 (MS and AV) and DE-FG02-11ER41742 Early Career Award (AV); the National Science Foundation under grant PHY-0643150 PECASE (AV); and the Sloan Research Foundation (AV).


\begin{thebibliography}{99}

\bibitem{Brink:1976bc}
  L.~Brink, J.~H.~Schwarz and J.~Scherk,
  ``Supersymmetric Yang-Mills Theories,''
  Nucl.\ Phys.\  B {\bf 121}, 77 (1977).

\bibitem{Aharony:1999ti}
  O.~Aharony, S.~S.~Gubser, J.~M.~Maldacena, H.~Ooguri and Y.~Oz,
  ``Large $N$ field theories, string theory and gravity,''
  Phys.\ Rept.\  {\bf 323}, 183 (2000)
  [arXiv:hep-th/9905111].

\bibitem{INTReview}
  N.~Beisert et al.,
  ``Special Volume: Review of AdS/CFT Integrability,''
  Lett.\ Math.\ Phys.\ {\bf 99} (2012) Numbers 1-3,
  [arXiv:1012.3982 [hep-th]].

\bibitem{SCATReview}
  R.~Roiban, M.~Spradlin and A.~Volovich (Eds.),
  ``Special issue: Scattering amplitudes in gauge theories:
  progress and outlook,''
  J.\ Phys.\ A {\bf 44} (2011) Number 45.

\bibitem{Witten:1998qj}
  E.~Witten,
  ``Anti-de Sitter space and holography,''
  Adv.\ Theor.\ Math.\ Phys.\  {\bf 2}, 253 (1998)
  [arXiv:hep-th/9802150].

\bibitem{Fitzpatrick:2011ia}
  A.~L.~Fitzpatrick, J.~Kaplan, J.~Penedones, S.~Raju and B.~C.~van Rees,
  ``A Natural Language for AdS/CFT Correlators,''
  JHEP {\bf 1111}, 095 (2011)
  [arXiv:1107.1499 [hep-th]].

\bibitem{Mack:2009mi}
  G.~Mack,
  ``$D$-independent representation of Conformal Field Theories in $D$
  dimensions
  via transformation to auxiliary Dual Resonance Models. Scalar amplitudes,''
  arXiv:0907.2407 [hep-th].

\bibitem{Penedones:2010ue}
  J.~Penedones,
  ``Writing CFT correlation functions as AdS scattering amplitudes,''
  JHEP {\bf 1103}, 025 (2011)
  [arXiv:1011.1485 [hep-th]].

\bibitem{Paulos:2011ie}
  M.~F.~Paulos,
  ``Towards Feynman rules for Mellin amplitudes in AdS/CFT,''
  JHEP {\bf 1110}, 074 (2011)
  [arXiv:1107.1504 [hep-th]].

\bibitem{Nandan:2011wc}
  D.~Nandan, A.~Volovich and C.~Wen,
  ``On Feynman rules for Mellin amplitudes in AdS/CFT,''
  arXiv:1112.0305 [hep-th].

\bibitem{Fitzpatrick:2011dm}
  A.~L.~Fitzpatrick and J.~Kaplan,
  ``Unitarity and the Holographic S-Matrix,''
  arXiv:1112.4845 [hep-th].

\bibitem{Polchinski:1999ry}
  J.~Polchinski,
  ``S-matrices from AdS spacetime,''
  arXiv:hep-th/9901076.

\bibitem{Susskind:1998vk}
  L.~Susskind,
  ``Holography in the flat space limit,''
  arXiv:hep-th/9901079.

\bibitem{Giddings:1999qu}
  S.~B.~Giddings,
  ``The boundary S-matrix and the AdS to CFT dictionary,''
  Phys.\ Rev.\ Lett.\  {\bf 83}, 2707 (1999)
  [arXiv:hep-th/9903048].

\bibitem{Okuda:2010ym}
  T.~Okuda and J.~Penedones,
  ``String scattering in flat space and a scaling limit of Yang-Mills
  correlators,''
  Phys.\ Rev.\  D {\bf 83}, 086001 (2011)
  [arXiv:1002.2641 [hep-th]].

\bibitem{Fitzpatrick:2011hu}
  A.~L.~Fitzpatrick and J.~Kaplan,
  ``Analyticity and the Holographic S-Matrix,''
  arXiv:1111.6972 [hep-th].

\bibitem{Raju:2012zr}
  S.~Raju,
  ``New Recursion Relations and a Flat Space Limit for AdS/CFT Correlators,''
  arXiv:1201.6449 [hep-th].

\bibitem{Alday:2007hr}
  L.~F.~Alday and J.~M.~Maldacena,
  ``Gluon scattering amplitudes at strong coupling,''
  JHEP {\bf 0706}, 064 (2007)
  [arXiv:0705.0303 [hep-th]].

\bibitem{Drummond:2007cf}
  J.~M.~Drummond, J.~Henn, G.~P.~Korchemsky and E.~Sokatchev,
  ``On planar gluon amplitudes/Wilson loops duality,''
  Nucl.\ Phys.\  B {\bf 795}, 52 (2008)
  [arXiv:0709.2368 [hep-th]].

\bibitem{Drummond:2007au}
  J.~M.~Drummond, J.~Henn, G.~P.~Korchemsky and E.~Sokatchev,
  ``Conformal Ward identities for Wilson loops and a test of the duality with
  gluon amplitudes,''
  Nucl.\ Phys.\  B {\bf 826}, 337 (2010)
  [arXiv:0712.1223 [hep-th]].

\bibitem{Drummond:2008vq}
  J.~M.~Drummond, J.~Henn, G.~P.~Korchemsky and E.~Sokatchev,
  ``Dual superconformal symmetry of scattering amplitudes in $\mathcal{N}=4$
  super-Yang-Mills theory,''
  Nucl.\ Phys.\  B {\bf 828}, 317 (2010)
  [arXiv:0807.1095 [hep-th]].

\bibitem{Drummond:2006rz}
  J.~M.~Drummond, J.~Henn, V.~A.~Smirnov and E.~Sokatchev,
  ``Magic identities for conformal four-point integrals,''
  JHEP {\bf 0701}, 064 (2007)
  [arXiv:hep-th/0607160].

\bibitem{ArkaniHamed:2010gh}
  N.~Arkani-Hamed, J.~L.~Bourjaily, F.~Cachazo and J.~Trnka,
  ``Local Integrals for Planar Scattering Amplitudes,''
  arXiv:1012.6032 [hep-th].

\bibitem{SmirnovBook}
  V.~A.~Smirnov,
  ``Evaluating Feynman Integrals,''
  Springer tracts in modern physics,
  {\bf 211} (Springer, Berlin, Heidelberg, 2004).

\bibitem{Bern:1994zx}
  Z.~Bern, L.~J.~Dixon, D.~C.~Dunbar and D.~A.~Kosower,
  ``One loop $n$ point gauge theory amplitudes, unitarity and collinear limits,''
  Nucl.\ Phys.\  B {\bf 425}, 217 (1994)
  [arXiv:hep-ph/9403226].

\bibitem{Bern:1994cg}
  Z.~Bern, L.~J.~Dixon, D.~C.~Dunbar and D.~A.~Kosower,
  ``Fusing gauge theory tree amplitudes into loop amplitudes,''
  Nucl.\ Phys.\  B {\bf 435}, 59 (1995)
  [arXiv:hep-ph/9409265].

\bibitem{Bern:2007ct}
  Z.~Bern, J.~J.~M.~Carrasco, H.~Johansson and D.~A.~Kosower,
  ``Maximally supersymmetric planar Yang-Mills amplitudes at five loops,''
  Phys.\ Rev.\  D {\bf 76}, 125020 (2007)
  [arXiv:0705.1864 [hep-th]].

\bibitem{ArkaniHamed:2010kv}
  N.~Arkani-Hamed, J.~L.~Bourjaily, F.~Cachazo, S.~Caron-Huot and J.~Trnka,
  ``The All-Loop Integrand For Scattering Amplitudes in Planar $\mathcal{N}=4$
  SYM,''
  JHEP {\bf 1101}, 041 (2011)
  [arXiv:1008.2958 [hep-th]].

\bibitem{CaronHuot:2010zt}
  S.~Caron-Huot,
  ``Loops and trees,''
  JHEP {\bf 1105}, 080 (2011)
  [arXiv:1007.3224 [hep-ph]].

\bibitem{Boels:2010nw}
  R.~H.~Boels,
  ``On BCFW shifts of integrands and integrals,''
  JHEP {\bf 1011}, 113 (2010)
  [arXiv:1008.3101 [hep-th]].

\bibitem{Goncharov:1994}
  A.~B.~Goncharov
  ``Polylogarithms and Motivic Galois groups,''
  Proc.\ Symp.\ Pure Math.\ {\bf 55}, 43 (1994).

\bibitem{Goncharov:1998}
  A.~B.~Goncharov,
  ``Multiple polylogarithms, cyclotomy and modular complexes,''
  Math.\ Res.\ Lett.\ {\bf 5}, 497 (1998).

\bibitem{Goncharov:2002}
  A.~B.~Goncharov,
  ``Galois symmetries of fundamental groupoids and noncommutative geometry,''
  Duke Math J.~{\bf 128}, 209 (2005),
  [arXiv:math/0208144].

\bibitem{Goncharov:2010jf}
  A.~B.~Goncharov, M.~Spradlin, C.~Vergu and A.~Volovich,
  ``Classical Polylogarithms for Amplitudes and Wilson Loops,''
  Phys.\ Rev.\ Lett.\  {\bf 105}, 151605 (2010)
  [arXiv:1006.5703 [hep-th]].

\bibitem{Gaiotto:2011dt}
  D.~Gaiotto, J.~Maldacena, A.~Sever and P.~Vieira,
  ``Pulling the straps of polygons,''
  JHEP {\bf 1112}, 011 (2011)
  [arXiv:1102.0062 [hep-th]].

\bibitem{DelDuca:2011ne}
  V.~Del Duca, C.~Duhr, V.~A.~Smirnov, C.~Duhr and V.~A.~Smirnov,
  ``The massless hexagon integral in $D = 6$ dimensions,''
  Phys.\ Lett.\  B {\bf 703} (2011) 363
  [arXiv:1104.2781 [hep-th]].

\bibitem{Dixon:2011ng}
  L.~J.~Dixon, J.~M.~Drummond and J.~M.~Henn,
  ``The one-loop six-dimensional hexagon integral and its relation to MHV
  amplitudes in $\mathcal{N}=4$ SYM,''
  JHEP {\bf 1106}, 100 (2011)
  [arXiv:1104.2787 [hep-th]].

\bibitem{DelDuca:2011jm}
  V.~Del Duca, C.~Duhr and V.~A.~Smirnov,
  ``The One-Loop One-Mass Hexagon Integral in $D=6$ Dimensions,''
  JHEP {\bf 1107}, 064 (2011)
  [arXiv:1105.1333 [hep-th]].

\bibitem{DelDuca:2011wh}
  V.~Del Duca, L.~J.~Dixon, J.~M.~Drummond, C.~Duhr,
  J.~M.~Henn and V.~A.~Smirnov,
  ``The one-loop six-dimensional hexagon integral with three massive corners,''
  Phys.\ Rev.\  D {\bf 84}, 045017 (2011)
  [arXiv:1105.2011 [hep-th]].

\bibitem{Spradlin:2011wp}
  M.~Spradlin and A.~Volovich,
  ``Symbols of One-Loop Integrals From Mixed Tate Motives,''
  arXiv:1105.2024 [hep-th].

\bibitem{Dixon:2011pw}
  L.~J.~Dixon, J.~M.~Drummond and J.~M.~Henn,
  ``Bootstrapping the three-loop hexagon,''
  JHEP {\bf 1111}, 023 (2011)
  [arXiv:1108.4461 [hep-th]].

\bibitem{Heslop:2011hv}
  P.~Heslop and V.~V.~Khoze,
  ``Wilson Loops @ 3-Loops in Special Kinematics,''
  JHEP {\bf 1111}, 152 (2011)
  [arXiv:1109.0058 [hep-th]].

\bibitem{Duhr:2011zq}
  C.~Duhr, H.~Gangl and J.~R.~Rhodes,
  arXiv:1110.0458 [math-ph].

\bibitem{Dixon:2011nj}
  L.~J.~Dixon, J.~M.~Drummond and J.~M.~Henn,
  ``Analytic result for the two-loop six-point NMHV amplitude in
  $\mathcal{N}=4$ super
  Yang-Mills theory,''
  JHEP {\bf 1201}, 024 (2012)
  [arXiv:1111.1704 [hep-th]].

\bibitem{CaronHuot:2011kk}
  S.~Caron-Huot and S.~He,
  ``Jumpstarting
  the all-loop S-matrix of planar $\mathcal{N}=4$ super Yang-Mills,''
  arXiv:1112.1060 [hep-th].

\bibitem{Prygarin:2011gd}
  A.~Prygarin, M.~Spradlin, C.~Vergu and A.~Volovich,
  ``All Two-Loop MHV Amplitudes in Multi-Regge Kinematics From Applied
  Symbology,''
  arXiv:1112.6365 [hep-th].

\bibitem{Brandhuber:2012vm}
  A.~Brandhuber, G.~Travaglini and G.~Yang,
  ``Analytic two-loop form factors in $\mathcal{N}=4$ SYM,''
  arXiv:1201.4170 [hep-th].

\bibitem{Duhr:2012fh}
  C.~Duhr,
  ``Hopf algebras, coproducts and symbols: an application to Higgs boson
  amplitudes,''
  arXiv:1203.0454 [hep-ph].

\bibitem{Laporta:2004rb}
  S.~Laporta and E.~Remiddi,
  ``Analytic treatment of the two loop equal mass sunrise graph,''
  Nucl.\ Phys.\  B {\bf 704}, 349 (2005)
  [arXiv:hep-ph/0406160].

\bibitem{Drummond:2010cz}
  J.~M.~Drummond, J.~M.~Henn and J.~Trnka,
  ``New differential equations for on-shell loop integrals,''
  JHEP {\bf 1104}, 083 (2011)
  [arXiv:1010.3679 [hep-th]].

\bibitem{Usyukina:1992jd}
  N.~I.~Usyukina and A.~I.~Davydychev,
  ``An Approach to the evaluation of three and four point ladder diagrams,''
  Phys.\ Lett.\  B {\bf 298}, 363 (1993).

\bibitem{Drummond:2010mb}
  J.~M.~Drummond and J.~M.~Henn,
  ``Simple loop integrals and amplitudes in $\mathcal{N}=4$ SYM,''
  JHEP {\bf 1105}, 105 (2011)
  [arXiv:1008.2965 [hep-th]].

\bibitem{Hodges:2009hk}
  A.~Hodges,
  ``Eliminating spurious poles from gauge-theoretic amplitudes,''
  arXiv:0905.1473 [hep-th].

\bibitem{Dirac:1936fq}
  P.~A.~M.~Dirac,
  ``Wave equations in conformal space,''
  Annals Math.\  {\bf 37}, 429 (1936).

\bibitem{Weinberg:2010fx}
  S.~Weinberg,
  ``Six-dimensional Methods for Four-dimensional Conformal Field Theories,''
  Phys.\ Rev.\  D {\bf 82}, 045031 (2010)
  [arXiv:1006.3480 [hep-th]].

\bibitem{Costa:2011dw}
  M.~S.~Costa, J.~Penedones, D.~Poland and S.~Rychkov,
  ``Spinning Conformal Blocks,''
  JHEP {\bf 1111}, 154 (2011)
  [arXiv:1109.6321 [hep-th]].

\bibitem{Costa:2011mg}
  M.~S.~Costa, J.~Penedones, D.~Poland and S.~Rychkov,
  ``Spinning Conformal Correlators,''
  JHEP {\bf 1111}, 071 (2011)
  [arXiv:1107.3554 [hep-th]].

\bibitem{Symanzik:1972wj}
  K.~Symanzik,
  ``On Calculations in conformal invariant field theories,''
  Lett.\ Nuovo Cim.\  {\bf 3}, 734 (1972).

\bibitem{ElShowk:2011ag}
  S.~El-Showk and K.~Papadodimas,
  ``Emergent Spacetime and Holographic CFTs,''
  arXiv:1101.4163 [hep-th].

\bibitem{Bern:1993kr}
  Z.~Bern, L.~J.~Dixon and D.~A.~Kosower,
  ``Dimensionally regulated pentagon integrals,''
  Nucl.\ Phys.\  B {\bf 412}, 751 (1994)
  [arXiv:hep-ph/9306240].

\end{thebibliography}
\end{document}